\documentclass[aps,reprint,nofootinbib,superscriptaddress]{revtex4-2}

\usepackage{amsmath,amsfonts,amssymb,amsthm,
                    mathtools, 
					  esint, 
				    hyperref, 
			      mathrsfs, 
		        graphicx} 

\usepackage[nodisplayskipstretch]{setspace}
\usepackage{acronym}
\usepackage[utf8]{inputenc}
\usepackage[T1]{fontenc}
\usepackage{graphicx}
\usepackage[version=4]{mhchem}
\usepackage{stmaryrd}
\usepackage{amsbsy}
\usepackage{latexsym}
\usepackage{comment}
\usepackage{natbib}
\usepackage{bm}
\usepackage{subfigure} 
\usepackage{color}
\usepackage{wasysym}
\usepackage{mathbbol}
\usepackage{bigints}
\allowdisplaybreaks
\usepackage[normalem]{ulem}
\usepackage[dvipsnames]{xcolor}
\usepackage{multirow}
\usepackage{physics}
\usepackage{siunitx}
\usepackage[export]{adjustbox}
\graphicspath{ {./images/} }
\usepackage{hyperref,url}
\hypersetup{
	colorlinks=true,       
	linkcolor=blue,          
	citecolor=blue,        
	urlcolor=blue           
}

\begin{document}
\title{Primordial Black Holes: A Review of Formation and Evolution}
\author{S. Shankaranarayanan}
\email{shanki@iitb.ac.in}
\author{Soumya Bhattacharya}
\email{soumya557@gmail.com}
\author{Archit Vidyarthi}
\email{architmedes@gmail.com}
\date{\today}
\affiliation{Department of Physics,  Indian Institute of Technology Bombay, Mumbai 400076, India}
\begin{abstract}
\noindent 
Primordial Black Holes (PBHs) have emerged as a leading non-particulate candidate for dark matter and a unique cosmological probe, a paradigm shift accelerated by the detection of anomalous binary mergers by the LIGO-Virgo-KAGRA (LVK) collaboration. 
While the literature is rich with phenomenological constraints, the fundamental quantum and relativistic underpinnings governing PBH genesis and evolution often receive comparatively less emphasis.
This review aims to bridge that gap by systematically detailing the physics of PBH formation and their subsequent evolutionary trajectory. We critically examine the hydrodynamic complexity of the early universe, establishing the relativistic thresholds for collapse, the non-linear \emph{race against sound} in the primordial plasma, and the rigorous mathematical utility of the compaction function. 
Furthermore, by incorporating the dynamic nature of FLRW backgrounds, higher curvature corrections, and quantum backreaction via the memory burden effect, we challenge the standard hawking evaporation and show that extreme-curvature environments halt evaporation entirely, leaving Planck-scale relics that evade current extragalactic bounds.
Finally, we map the multimessenger observational landscape, highlighting how the imminent search for sub-solar mass inspirals by next-generation gravitational wave observatories --- such as the Einstein Telescope and Cosmic Explorer --- could yield smoking-gun evidence for the PBH paradigm, ultimately transforming these primordial relics into unparalleled laboratories for high-energy physics.
\end{abstract}
\maketitle
\acrodef{DM}[DM]{Dark Matter}
\acrodef{EMD}[EMD]{Early Matter-Dominated}
\acrodef{RD}[RD]{Radiation-Dominated}
\acrodef{GR}[GR]{General Relativity}
\acrodef{BH}[BH]{black hole}
\acrodefplural{BH}[BHs]{Black Holes}
\acrodef{PBH}[PBH]{Primordial Black Hole}
\acrodefplural{PBH}[PBHs]{Primordial Black Holes}
\acrodef{SMBH}[SMBH]{supermassive black hole}
\acrodef{LVK}[LVK]{LIGO-Virgo-Kagra Collaborations}
\acrodef{GW}[GW]{Gravitational Wave}
\acrodefplural{GW}[GWs]{gravitational waves}
\acrodef{NR}[NR]{Numerical Relativity}
\acrodef{SIGWs}[SIGWs]{Second Order Induced GWs}
\acrodef{PTA}[PTA]{Pulsar Timing Arrays}
\acrodef{LSS}[LSS]{Large Scale Structure}
\acrodef{CMBR}[CMBR]{Cosmic Microwave Background Radiation}
\acrodef{FLRW}[FLRW]{Friedman-Lemaître-Robertson-Walker}
\acrodef{CDM}[CDM]{Cold Dark Matter}
\acrodef{MTS}[MTS]{Marginally Trapped Surface}
\section{Introduction} 

The nature of \ac{DM} remains one of the most persistent anomalies in modern cosmology~\cite{Bertone:2016nfn}. While the particle physics community has long favored Weakly Interacting Massive Particles (WIMPs), the enduring absence of detection in direct capture experiments and collider searches has forced widening of the theoretical horizon~\cite{Roszkowski:2017nbc,Billard:2021uyg,Kahn:2021ttr,PerezdelosHeros:2020qyt,Misiaszek:2023sxe}. In this context, \acp{PBH} --- macroscopic objects formed in the early universe through the gravitational collapse of extreme density fluctuations --- have re-emerged as a compelling non-particulate candidate~\cite{Zeldovich:1967lct,Hawking:1971ei}. Unlike stellar \acp{BH}, which mark the endpoint of stellar evolution, \acp{PBH} are relics of the early
universe~\cite{Carr:1974nx,Carr:1975qj,Nadezhin:1978aa}. 

PBHs occupy a \emph{unique theoretical niche}: they are the only \ac{DM} candidate that \emph{does not} strictly require new particle species~\cite{Zeldovich:1967lct,Hawking:1971ei}. 
While their formation typically demands an enhancement of the primordial power spectrum~\cite{Carr:2021bzv} --- often sourced by specific inflationary dynamics --- these compact objects can be assumed to be composed of Standard Model particles that have undergone gravitational collapse, effectively providing a macroscopic dark matter candidate without invoking new physics~\cite{Khlopov:2008qy,Carr:2021bzv,Carr:2026hot,Green:2020jor,Escriva:2022duf,Choudhury:2024aji}.
From a dynamical perspective, they behave as \ac{CDM} --- non-relativistic, effectively collisionless, and interacting solely via gravity.

Unlike particle colliders, which are limited by terrestrial energy constraints ($\sim 14$ TeV), PBHs serve as a unique probe of the high-energy Universe. Their formation connects the non-linear dynamics of \ac{GR}~\cite{Musco:2018rwt,Escriva:2019phb} directly to the quantum fluctuations of inflation at extremely small scales ($k \sim 10^{12-18} \text{Mpc}^{-1}$)~\cite{Green:1997sz,Sasaki:2018dmp}. Consequently, the PBH mass spectrum acts as a fossilized record of the primordial power spectrum and the thermal history of the Universe, providing us access to physical regimes otherwise lost to observation~\cite{Carr:2020xqk,Byrnes:2018clq,Young:2019yug}.

The discovery of \acp{GW} by the \ac{LVK} has revitalized the PBH hypothesis~\cite{Sasaki:2016jop,Bird:2016dcv,Clesse:2016vqa}. The inaugural detection, GW150914, revealed a binary merger of $\sim 30 M_{\odot}$ \acp{BH} --- a mass scale unexpectedly high for standard stellar remnants~\cite{LIGOScientific:2016aoc}. Subsequent detections have deepened the intrigue: First, events like GW190521 involve progenitors in the \emph{upper mass gap} ($\sim 60-120 M_{\odot}$), a region forbidden to stellar \acp{BH} due to pair-instability supernovae. Second, the characteristically low effective inspiral spins ($\chi_{\text{eff}} \approx 0$) observed in many \ac{LVK} events~\cite{Fernandez:2019kyb,Franciolini:2021tla} align naturally with PBH formation scenarios, whereas stellar binary evolution often predicts higher spin retention~\cite{Bavera:2020inc,Gerosa:2021mno}. Table \eqref{tab:GWEvents} lists the key GW events supporting the PBH hypothesis.
\begin{figure*}[!htb]
\centering
\includegraphics[width=1.0\textwidth]{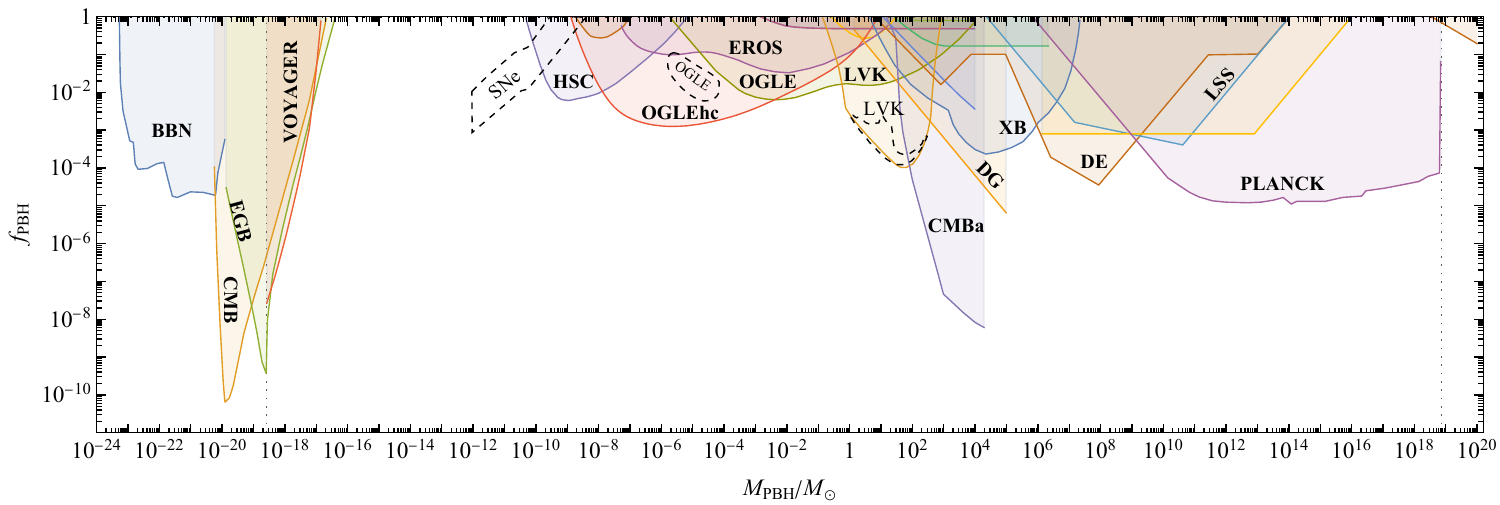} 
\caption{The Constraint \emph{reprinted plot}  for PBH Abundance (assuming monochromatic PBH mass function). This plot illustrates the maximum allowed fraction of dark matter in the form of PBHs, $f_{\rm PBH} \equiv \Omega_{\rm PBH}/\Omega_{\rm DM}$, as a function of the PBH mass $M$ in units of solar mass  ($M_{\odot}$). Reprinted from, Carr et al. (2026)~\cite{Carr:2026hot}.}
\label{fig:MasterConstraint}
\end{figure*}
\begin{table*}[!htb]
\centering\caption{Key GW events supporting the PBH hypothesis}
\label{tab:GWEvents}
\begin{tabular}{l c l}\hline\textbf{Event} & \textbf{Masses ($M_{\odot}$)} & \textbf{Significance for PBH Scenarios} \\ \hline
GW150914~\cite{LIGOScientific:2016aoc} & $36, 29$ & Unexpectedly heavy for stellar remnants; revived PBH interest. \\
GW190814~\cite{LIGOScientific:2020zkf} & $23, 2.6$ & Secondary component falls in \emph{lower mass gap} (too heavy for Neutron Star, too light for BH) \\ 
GW190521~\cite{LIGOScientific:2020ufj} & $85, 66$ & Progenitors reside in the \emph{upper mass gap} forbidden by pair-instability supernovae. \\ 
GW250114~\cite{LIGOScientific:2025rid} & $36,30$ & \emph{Highest SNR ($\sim 80$) event till date}. \\ 
\hline
\end{tabular}
\end{table*}

The landmark detection of GW150914, alongside recent high signal-to-noise ratio (SNR) events involving the merger of $\sim 36 \, M_\odot$ and $\sim 32 \, M_\odot$ \acp{BH}~\cite{LIGOScientific:2025rid}, initially surprised the astrophysical community due to the unexpectedly large masses of the progenitors. While standard solar-metallicity stellar winds shed too much mass to leave behind such heavy remnants, subsequent binary population synthesis models have demonstrated that these $\sim 30 \, M_\odot$ \acp{BH} can successfully form via classical isolated binary evolution in low-metallicity environments ($Z \sim 0.001$)~\cite{Stevenson:2017tfq}. Nevertheless, while astrophysical channels can account for these specific masses, the overarching demographic anomalies in the growing \ac{GW} catalog --- such as near-zero effective inspiral spins and events falling squarely within the pulsational pair-instability mass gap ---strongly motivate the \ac{PBH} paradigm as a compelling concurrent population.

Fig.~\ref{fig:MasterConstraint} summarizes the current observational landscape for PBH dark matter across varying mass scales. The plot is a summary of exclusion limits spanning nearly twenty orders of magnitude in mass and serves a dual purpose: it identifies the narrow \emph{open windows} where PBHs may constitute the totality of \ac{DM} ($f_{\text{PBH}} \approx 1$), while simultaneously demarcating the regimes where they are restricted to a sub-dominant fraction. 

In the macroscopic regime ($10^{-11} M_{\odot} \lesssim M \lesssim 10^{2} M_{\odot}$), the parameter space is tightly bound by various gravitational microlensing surveys --- including Subaru Hyper Suprime-Cam (HSC), Optical Gravitational Lensing Experiment (OGLE), EROS, and MACHO --- which severely restrict the allowed PBH abundance~\cite{Niikura:2017zjd, Tisserand:2006zx, Allsman:2000kg, Mroz:2024mse, Mroz:2024wia}. Below this, we encounter the so-called \emph{Asteroid-Mass Window} ($10^{-16} M_{\odot} \lesssim M \lesssim 10^{-11} M_{\odot}$), which currently represents the primary unconstrained parameter space where PBHs could potentially account for the entirety of \ac{DM} ($f_{\text{PBH}} = 1$). In this regime, traditional optical microlensing loses efficacy due to the breakdown of the geometric optics approximation and finite-source-size effects~\cite{Katz:2018zrn, Smyth:2019whb}. At the heavier end of the spectrum ($M \gtrsim 10 M_{\odot}$), the abundance is heavily constrained by \ac{LVK} binary merger rates as well as the thermal signatures of cosmological gas accretion imprinted on the \ac{CMBR}~\cite{Ali-Haimoud:2016mbv, LIGOScientific:2019kan}. Finally, at the highest masses ($M \gtrsim 10^{3} M_{\odot}$), the limits are driven by large-scale astrophysical dynamics, including dynamical friction, the disruption of wide binaries, and modifications to \ac{LSS} via Poissonian fluctuations~\cite{Carr:2020gox}.

Thus, within this landscape, two regions are of particular phenomenological importance. The \emph{asteroid-mass window} ($10^{17}$–$10^{23}$ g) represents the primary unconstrained parameter space for $f_{\text{PBH}} = 1$~\cite{Niikura:2017zjd,Tinyakov:2024mcy}. This window is bounded at lower masses by the non-observation of Hawking evaporation products (e.g., extragalactic gamma-ray background constraints from Fermi-LAT) and at higher masses by microlensing limits (HSC/Subaru and OGLE). Notably, microlensing constraints loosen at the asteroid scale due to finite-source-size effects and wave optics interference, which arise when the PBH Schwarzschild radius becomes comparable to the wavelength of the background source light~\cite{Niikura:2017zjd, Tisserand:2006zx, Allsman:2000kg}.

Conversely, the \emph{Solar and Sub-Solar Mass Window} ($\sim 0.1$–$100~M_{\odot}$) represents a region of intense activity driven by \ac{GW} astronomy. While the LVK collaboration has detected mergers in the $10$–$100 M_{\odot}$ range, theoretical interest is increasingly shifting toward the \emph{sub-solar} regime ($\sim 0.1 M_{\odot}$), which offers a unique multiband test of the PBH hypothesis. A population of sub-solar PBHs provides a distinct advantage: it can be probed simultaneously by \ac{PTA} at formation and ground-based interferometers during merger. We can verify this via a simple order-of-magnitude estimate. The frequency of standard \ac{SIGWs} associated with the formation of a PBH of mass $M$ scales as~\cite{Wang:2019kaf}:
\begin{equation}
f_{\text{form}} \approx 1.5 \times 10^{-9}  \left( \frac{M}{30 M_{\odot}} \right)^{-1/2} \, 
\text{ Hz} \, . 
\end{equation}
For a sub-solar mass $M \sim 0.1 M_{\odot}$, the formation signal peaks at $f_{\text{form}} \sim 2 \times 10^{-8}$ Hz, falling squarely within the sensitivity band of PTAs like NANOGrav and SKAO~\cite{NANOGrav:2023hvm,Santos:2025fff}. 

However, these same objects will eventually form binaries and merge. The frequency of the GWs emitted during the late inspiral and merger is bounded by the Innermost Stable Circular Orbit (ISCO)~\cite{Sasaki:2018dmp}:
\begin{equation}
f_{\text{merger}} \approx \frac{1}{6\sqrt{6}\pi} \frac{c^3}{G M} \approx 4.4  \left( \frac{M_{\odot}}{M_{\rm Tot}} \right) \text{ kHz} \, .
\end{equation}
where $M_{\rm Tot} = m_1+m_2$ is the total mass of the PBH binaries with individual masses $m_1$ and $m_2$. 
For $M_{\rm Tot} \sim 0.1 M_{\odot}$, the merger frequency is $f_{\text{merger}} \sim 44$ kHz. Consequently, the inspiral phase sweeps through the $10$--$1000$ Hz band, making it detectable by current \ac{LVK} detectors and future third-generation experiments like the Cosmic Explorer (CE), Einstein Telescope (ET)~\cite{Dandoy:2023jot,LIGOScientific:2021job}. Thus, the sub-solar window allows for a powerful consistency check: the stochastic background from their birth must be visible in PTAs, while their discrete mergers must appear in high-frequency GW detectors~\cite{Aggarwal:2025noe}.

The literature is currently rich with comprehensive reviews detailing the observational constraints on PBH abundance, covering everything from microlensing and \ac{CMBR} distortions to dynamical friction \cite{Carr:2020xqk,Carr:2021bzv,Green:2020jor,Escriva:2022duf,Carr:2020gox}. However, for researchers in gravitational physics and high-energy theory, the fundamental physical mechanisms governing PBH genesis and quantum evolution are often obscured by the focus on phenomenological bounds. \emph{This review aims to bridge that gap}. 
Rather than focusing primarily on where to find them, we focus on how they form and evolve. We systematically derive the relativistic thresholds for collapse, the hydrodynamic \emph{race against sound} in the primordial plasma, and the quantum mechanical decay that dictates their ultimate fate.

Furthermore, a rigorous understanding of these formation mechanics transforms PBHs into a unique diagnostic tool for the Early Universe. While the \ac{CMBR} and \ac{LSS} precisely constrain the inflationary power spectrum at large scales ($k \lesssim 0.1 \, \text{Mpc}^{-1}$), the dynamics of the final $\sim 30$ e-folds of inflation remain largely unprobed~\cite{Gow:2020bzo}. 
PBHs act as a \emph{cosmological laboratory} for this regime: their abundance depends exponentially on the amplitude of primordial fluctuations at small scales ($k \sim 10^{5}$--$10^{23} \, \text{Mpc}^{-1}$)\footnote{In the literature, it is widely assumed that the enhancement of the primordial power spectrum ($P_{\zeta} \sim 10^{-2}$) is an outcome of inflationary dynamics, often achieved via an ultra-slow-roll (USR) phase~\cite{Garcia-Bellido:2017mdw}.}~\cite{Motohashi:2017kbs}. Consequently, the detection (or non-detection) of PBHs --- and the stochastic background of \ac{SIGWs} inevitably generated during their formation --- provides the only direct window into the physics of the late inflationary epoch and the reheating transition~\cite{Inomata:2016rbd,Garcia-Bellido:2017aan,Sasaki:2018dmp,Domenech:2021ztg}.

\begin{figure}[h!]
\centering
\includegraphics[width=0.5\textwidth]{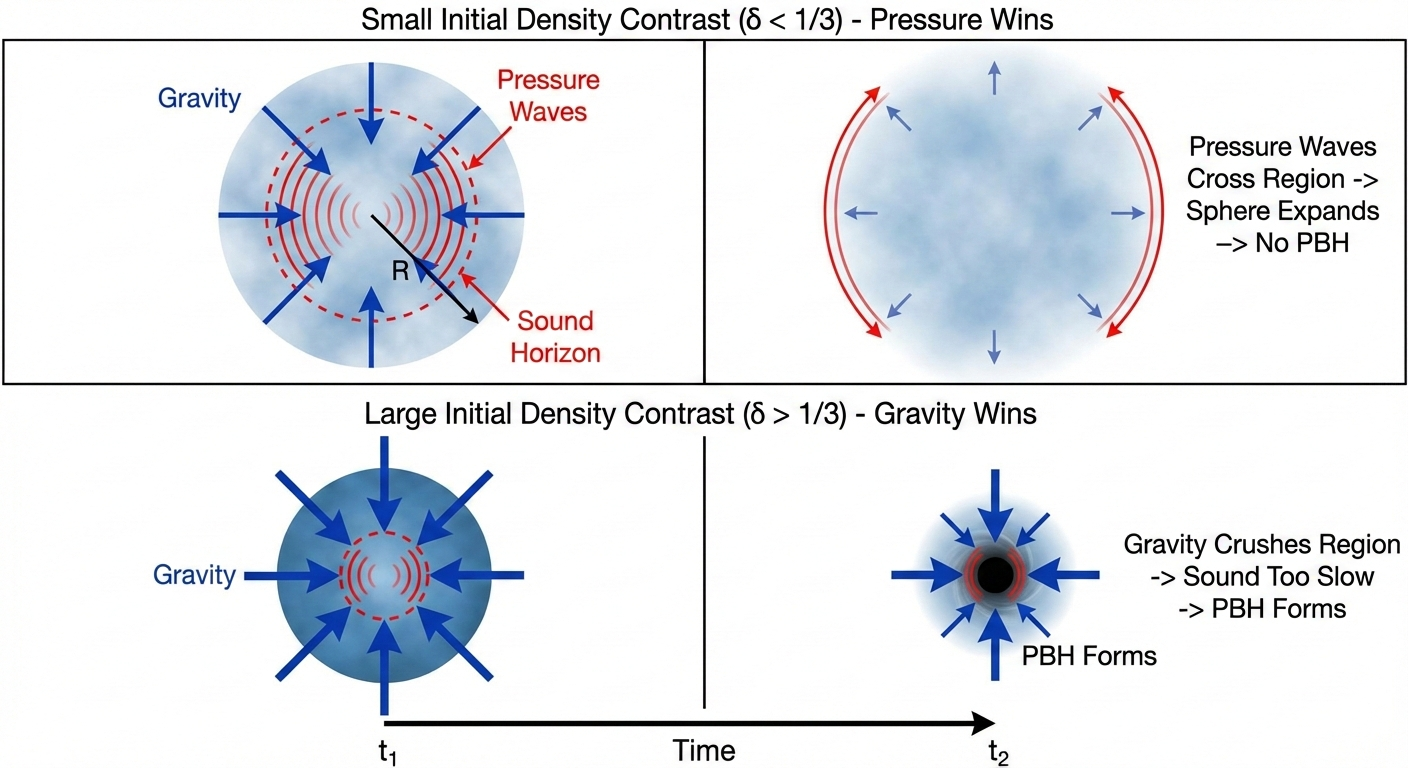} 
\caption{Illustration of the PBH formation in the rare overdense region in the early universe. Credit: Nanobanana}
\label{fig:PBHformation}
\end{figure}

A key challenge in PBH theory lies in accurately modeling the hydrodynamic complexity of formation. The standard scenario assumes the collapse of Hubble-sized overdensities during the \ac{RD} era. However, this is not a simple free-fall. As illustrated in Fig.~\eqref{fig:PBHformation}, when a perturbation enters the horizon, it must overcome the immense radiation pressure ($P = \rho c^2/3$) of the early universe. While simple analytic estimates based on the Jeans criterion suggest a fixed threshold proportional to the sound speed squared ($\delta_c \approx c_s^2 = c^2/3$), full general relativistic hydrodynamics simulations reveal a more complex reality. Numerical studies demonstrate that $\delta_c$ is not a universal constant but a dynamical variable capable of ranging between $0.4$ and $0.66$, depending sensitively on the specific shape of the curvature profile and transfer function used to profile the initial density perturbations~\cite{Musco:2004ak,Musco:2018rwt,Escriva:2019phb}. 

It is crucial to emphasize that the formation of a \ac{PBH} is an exceptionally rare event. Current observational constraints require the initial mass fraction at formation to be of the order $\beta \sim 10^{-9}$ for solar-mass PBHs (see the discussion in Sec.~\eqref{sec:AbundanceDerivation}), and as low as $\sim 10^{-15}$ for asteroid-mass candidates. Physically, this implies that for every billion to quadrillion Hubble patches in the early universe, only one succeeded in overcoming pressure forces to collapse; the rest simply dispersed as acoustic waves. If this formation rate were even slightly higher, the subsequent growth of the PBH density contrast (which scales as $a(t)$ relative to radiation) would have led to a \emph{PBH-dominated} universe long before the onset of structure formation --- a scenario strictly ruled out by observations~\cite{Carr:1975qj,Green:2020jor,Carr:2009jm}.
Lastly, since PBHs form soon after horizon re-entry in the \ac{RD} era, their mass is approximately in one-to-one correspondence with the scale at which the primordial curvature perturbations peak. This relation enables a direct translation of constraints: for example, bounds on PBH abundance can be used to constrain the primordial curvature power spectrum (see Ref.~\cite{Kushwaha:2026msi} for updated results, along with \cite{Sato-Polito:2019hws,Kalaja:2019uju,Gow:2020bzo}), as well as the power spectrum of primordial magnetic fields during the RD epoch~\cite{Kushwaha:2024zhd}.
Also, for completeness, we contrast this radiative collapse with formation during an \ac{EMD} era. In regimes where the pressure vanishes (e.g., during reheating or phase transitions), the Jeans length drops to zero. Consequently, the \emph{race against sound} described above does not apply; formation can occur on sub-horizon scales, and the primary barrier to collapse shifts from pressure gradients to deviations from sphericity and angular momentum~\cite{Khlopov:1980mg,Harada:2016mhb}.

Finally, we extend the discussion to the quantum limit. While semi-classical theory predicts that PBHs lighter than $10^{15}$ g should have evaporated by today, recent developments in quantum information storage --- specifically the \emph{memory burden} effect~\cite{Dvali:2018xpy,Dvali:2021tez} --- suggest that backreaction may halt evaporation. This introduces the tantalizing possibility that Planck-scale remnants of primordial black holes could persist as stable \ac{DM} relics, radically altering the constraints on the low-mass window.

Before proceeding, it is necessary to explicitly define the physical scope and baseline assumptions of this review. Unless otherwise specified, our analytical framework assumes that \acp{PBH} form from primordial density perturbations governed by Gaussian statistics, and that the subsequent gravitational collapse occurs within spherically symmetric overdense regions. While primordial non-Gaussianities and anisotropic collapse dynamics --- such as triaxial or spindle collapse during an \ac{EMD} epoch --- can substantially alter formation thresholds, mass functions, and abundance predictions, they are treated herein as extensions to this standard paradigm. In sections where the underlying physical mechanisms necessitate relaxing these strict baseline assumptions, the deviations will be explicitly mentioned.

The remainder of this review is organized as follows. Sections \eqref{sec:SphSy} and \eqref{sec:Uniqueness} establish the geometric foundations of spherically symmetric black holes and the classical uniqueness theorems. Sections \eqref{sec:CosmoBH} through \eqref{sec:Compaction} transition to the dynamic early universe, systematically detailing the relativistic hydrodynamics of PBH formation. Here, we analyze the non-linear \emph{race against sound} within the primordial plasma and rigorously define the threshold for gravitational collapse using the compaction function. Section \eqref{sec:Abundance} bridges these local formation mechanics to global cosmology, utilizing statistical frameworks to quantify the initial PBH mass fraction and their present-day \ac{DM} abundance. Following this, we explore the quantum mechanical evolution of PBHs; specifically, Section \eqref{sec:HawkingRad} critically examines the limitations of standard semiclassical Hawking evaporation, discussing how dynamic cosmological backgrounds, higher-curvature corrections, and the memory burden effect might allow low-mass PBHs to survive as stable relics. Finally, Section \eqref{sec:Conclusion} concludes with a multimessenger observational roadmap, highlighting how next-generation gravitational wave and electromagnetic observatories will definitively test the PBH paradigm. The five appendices contain the mathematical details. 

We work with the metric signature $(-,+,+,+)$ and set $\hbar = c = 1$. In Sec.~\eqref{sec:FormationMechanics} and Appendices we explicitly use the speed of light ($c$). 

\section{Spherically symmetric black hole in Asymptotically flat space-times}
\label{sec:SphSy}

We start with the general 4-dimensional, static, spherically symmetric line element given by:
\begin{eqnarray}
\label{eq:gen-4D}
ds^2 &=& - f(r) dt^2 + \frac{dr^2}{g(r)} + R^2(r) d\Omega^2 \, , \\ 
\label{eq:spher-tx}
&=&  f(r)\left[ -dt^2 + dx^2\right] + R^2(r) d\Omega^2 \, ,
\end{eqnarray}
where $f(r)$, $g(r)$ and $R(r)$ (areal radius) are arbitrary smooth functions of the radial coordinate $r$. Here, $d\Omega^2$ is the metric on the unit 2-sphere. The \emph{tortoise coordinate} $x$ is:
\begin{equation}
x = \int {dr }/{\sqrt{f(r) g(r)}} ~.
\label{eq:rela-xr}
\end{equation}
We assume the spacetime is asymptotically flat and contains an event horizon at $r = r_h$.

\subsection{Horizon structure and Surface Gravity}

For the static metric in Eq. (\ref{eq:gen-4D}), there exists a timelike Killing vector field $\xi^{\mu} = (\partial_t)^\mu = (1, 0,\cdots, 0)$. The surface gravity $\kappa$ is defined via $\xi^\nu \nabla_\nu \xi^\mu = \kappa \xi^\mu$ evaluated at the horizon~\cite{Wald:1984rg}. Using the formula $\kappa^2 = - \frac{1}{2} (\nabla^\mu \xi^\nu) (\nabla_\mu \xi_\nu)$, we obtain~\cite{Wald:1984rg}:
\begin{equation}
\kappa = \frac{1}{2} \left(\sqrt{\frac{g(r)}{f(r)}} \frac{df(r)}{dr}\right)_{r = r_h} \, .
\label{eq:kappa-def}
\end{equation}
The event horizon is a null hypersurface defined by $g^{\mu\nu} \partial_{\mu} r \partial_{\nu} r = g(r) = 0$. Since $f(r)$ and $g(r)$ typically vanish at the horizon, we define the ratio $H(r) = f(r)/g(r)$ to be regular and non-vanishing at $r_h$.

To analyze the geometry near the horizon, we transform to Rindler-like coordinates. Defining the proper distance $\gamma \equiv \int dr/\sqrt{g} \approx \frac{1}{\kappa}\sqrt{f}$ near $r_h$, the metric approaches:
\begin{equation}
ds^2 \to - \kappa^2 \gamma^2 dt^2 + d\gamma^2 + R^2(r_h) \, d\Omega^2 \, .
\end{equation}
This reveals that the near-horizon geometry is effectively Rindler space (flat space seen by an accelerated observer) cross a sphere.

The behavior of the tortoise coordinate $x$ (Eq. (\ref{eq:rela-xr})) depends on whether the horizon is degenerate (extremal) or non-degenerate.
For non-degenerate horizon, like Schwarzschild, the functions $f(r)$ and $g(r)$ have simple zeros at $r_h$: $f(r) \sim f'(r_h)(r-r_h)$. The integration yields a logarithmic divergence:
\begin{equation}
x \sim \frac{1}{2\kappa} \ln(r - r_h) \quad \text{as}~ r \to r_h.
\label{eq:rel-xrh}
\end{equation}
This allows the horizon to be pushed to $x \to -\infty$.

In the case of extremal \acp{BH}, like  Extremal Reissner-Nordstr\"om, $r_h$ is a double root, so $f(r) \sim (r-r_h)^2$. The integration yields an inverse power law:
\begin{equation}
x \sim - \frac{c_0}{r - r_h} \quad (\text{as } r \to r_h).
\end{equation}
This corresponds to an infinitely long \emph{throat region} which is physically distinct from the Schwarzschild case.

\subsection{Singularities and Curvature Scales}
While the event horizon ($r_h = 2GM/c^2$) is a coordinate singularity (removable by coordinate transformation), the center at $r=0$ represents a physical curvature singularity. To quantify the strength of gravity, we examine the Kretschmann scalar $K = R_{\mu\nu\rho\sigma}R^{\mu\nu\rho\sigma}$.
For a 4D Schwarzschild \ac{BH}, this invariant is:
\begin{equation}
K = \frac{48 G^2 M^2}{r^6} \, .
\end{equation}
Evaluating this at the event horizon  reveals a crucial scaling law for \acp{PBH} in-terms of the solar mass:
\begin{equation}
K(r_h) \propto \frac{1}{G^4 M^4} \approx 1.55 \times 10^{-13} \left( \frac{M}{M_\odot} \right)^{-4} \, \text{m}^{-4}  \, .
\end{equation}
This scaling carries important physical implications.
For astrophysical \acp{BH} formed via stellar collapse, the curvature at the horizon is exceedingly small, i. e., $K \sim 10^{-13} \mathrm{~m}^{-4}$, and allows for a strictly classical treatment. However, for \acp{PBH} in the so-called \emph{asteroid-mass window} ($10^{17} - 10^{23}$ g), the spacetime curvature at the horizon is enormous, i. e., $K \sim 10^{51} \mathrm{~m}^{-4}$~\cite{Shankaranarayanan:2022wbx,Mandal:2025xuc}. Such extreme curvature implies that higher-order gravitational corrections (e.g., $O\left(R^2\right)$ or Gauss-Bonnet terms) and quantum effects become dominant factors in the evolution of PBH. Because of this extreme curvature, these low-mass PBHs serve as unique, natural laboratories for testing quantum gravity and effective field theories. In this regime, the standard Einstein-Hilbert action is no longer sufficient; higher-order curvature corrections --- such as Gauss-Bonnet or $\mathcal{O}(R^2)$ terms—become non-negligible at the horizon itself, potentially modifying the BH's geometry and thermodynamic properties~\cite{Calmet:2014dea,Mandal:2023kpu}. 

Furthermore, this extreme environment necessitates the inclusion of dynamic quantum backreaction, challenging the assumption of a static, unperturbed geometry. A critical recent development in this context is the \emph{memory burden} effect~\cite{Dvali:2018xpy, Dvali:2021tez}. In standard semi-classical Hawking evaporation, a BH is assumed to lose mass smoothly until it disappears. However, the memory burden paradigm identifies a physical mechanism where the enormous quantum information (or "soft hair") stored by the BH creates a severe backreaction on the geometry~\cite{Parvez:2025wtq}. 

As the PBH evaporates and shrinks, it enters a highly degenerate quantum state. The backreaction from this stored information suppresses the rate of particle emission, significantly delaying the final stages of evaporation~\cite{Xavier:2021chn,Thoss:2024hsr}. Integrating this effect bridges the gap between the static geometric description of the horizon and its dynamic quantum evolution: it suggests that asteroid-mass PBHs might live orders of magnitude longer than standard semi-classical predictions, radically altering their viability as a \ac{DM} candidate~\cite{Parvez:2025wtq}.

\subsection{The Hoop Conjecture}
\label{sec:Hoop}

The Hoop Conjecture, originally formulated by Kip Thorne in 1972, acts as the fundamental geometric precursor to the modern criteria for \ac{PBH} formation. In asymptotically flat spacetime, the criteria for \ac{BH} formation is encapsulated by Thorne's \emph{Hoop Conjecture}~\cite{Thorne:1972abc}. It states that an imploding mass $M$ forms a \ac{BH} if and only if it gets compacted into a region whose circumference $C$ in every direction satisfies~\cite{bonnor1983hoop,flanagan1991hoop,Senovilla:2007dw,hod2018status}:
\begin{equation}
\label{eq:HoopConj01}
C \lesssim 4\pi G M \, .
\end{equation}
For a spherical object of radius $R_0$, this simplifies to the condition $R_0 \lesssim 2GM$, or equivalently, the compactness ratio $\mathcal{C}_{HC}$:
\begin{equation}
\mathcal{C}_{HC} \equiv \frac{2GM}{R_0} \gtrsim 1 \, .
\label{def:Compactness}
\end{equation}
This geometric intuition is fundamental. As we transition to the cosmological context in Sec. (\ref{sec:CosmoBH}), we will see that while the expansion of the Universe complicates the definition of Mass and Radius, the core idea of the Hoop Conjecture survives in the form of the \emph{Compaction Function} which is now considered the standard, rigorous quantity for evaluating \ac{PBH} formation criteria~\cite{Shibata:1999zs,Musco:2018rwt}.

However, one must exercise caution when comparing threshold values across the literature due to differing conventions. Shibata and Sasaki~\cite{Shibata:1999zs} originally introduced the compaction function purely as the ratio of the mass excess to the areal radius:
\begin{equation}
    \mathcal{C}_{\text{SS}}(r) = \frac{G\, \delta M(r)}{R(r)} \, .
\end{equation}
In contrast, many modern \ac{NR} studies (for instance, \cite{Musco:2018rwt}) include a factor of 2 to align the definition more explicitly with the Schwarzschild horizon limit (the Hoop Conjecture):
\begin{equation}
\label{def:Compaction}
    \mathcal{C}(r) = \frac{2G\, \delta M(r)}{R(r)} \, .
\end{equation}
While physically equivalent, this trivial factor of two leads to different numerical thresholds for collapse. 
Note that the modern convention is widely considered superior and is now the standard in \ac{NR} simulations~\cite{Musco:2018rwt,Escriva:2025rja}. By including the factor of 2, the modern definition elegantly maps the local collapse threshold directly to the Schwarzschild radius limit dictated by the Hoop Conjecture. Consequently, as we will see in Sec. \eqref{sec:DetailedCompaction}, when the peak of the compaction function approaches $\mathcal{C}_{\text{max}} \sim 0.4 - 0.5$ (in a \ac{RD} background), it intuitively signals to numerical relativists that the local overdensity is approaching the absolute $\mathcal{O}(1)$ threshold of horizon formation. This direct structural alignment with the Schwarzschild limit provides a mathematically robust and physically intuitive diagnostic for tracking the non-linear collapse of PBHs in \ac{NR} codes. (Details in Sec. \eqref{sec:DetailedCompaction}.)

\section{Uniqueness Theorems}
\label{sec:Uniqueness}

To understand why the Schwarzschild metric is the standard description for non-spinning \acp{BH}, we invoke two fundamental theorems in \ac{GR}.

\subsection{Birkhoff's Theorem}

This theorem states that any spherically symmetric solution to the vacuum Einstein field equations is necessarily static and asymptotically flat. The unique solution satisfying these conditions is the Schwarzschild metric \cite{1923-Birkhoff-book,2005-Jebsen-GRG}. However, this may be violated for modified gravity theories~\cite{Xavier:2020ulw}.

This theorem is powerful because it implies that a spherically symmetric vacuum region --- even if embedded in a dynamic or collapsing environment --- is described by the Schwarzschild geometry. However, it strictly applies only to vacuum; during the formation of \acp{PBH} within the dense radiation fluid of the early Universe, the interior metric is dynamic (e.g., \ac{FLRW}) until the collapse is complete and a vacuum horizon forms~\cite{Carr:1974nx,Polnarev:2006aa}. 

While Birkhoff's Theorem provides a powerful justification for describing isolated PBHs with the standard Schwarzschild (or Kerr) metric, it is crucial to recognize the limitations of this theorem in the extreme environment of the early universe. Birkhoff's Theorem strictly applies to spherically symmetric vacuum solutions in standard General Relativity. However, this assumption of a pristine vacuum and absolute staticity is frequently challenged by both the dynamic cosmological background and alternative theories of gravity~\cite{Xavier:2020ulw}.

In the context of early-universe phase transitions or modified gravity theories --- such as $f(R)$ gravity or Einstein-Dilaton-Gauss-Bonnet (EDGB) gravity --- the generalized versions of Birkhoff's Theorem often break down~\cite{Xavier:2020ulw}. In these frameworks, spherically symmetric solutions are not guaranteed to be static, nor are they restricted to the standard Schwarzschild geometry~\cite{Sotiriou:2013qea}. Furthermore, the primordial plasma is populated by dynamic scalar fields, such as the inflaton or curvaton. 

\subsection{The No-Hair Theorem}
While Birkhoff's theorem addresses spherical symmetry, the \emph{No-Hair Theorem} (or Conjecture) generalizes the uniqueness to rotating and charged bodies. It asserts that all stationary, asymptotically flat \ac{BH} solutions in \ac{GR} are uniquely characterized by  three externally observable classical parameters \cite{1967-Israel-PhyRev,1971-Carter-PRL,Hawking:1972qk} --- 
Mass ($M$), Electric Charge ($Q$) and Angular Momentum ($J$). Essentially, all other information (``hair") about the matter that formed the \ac{BH} --- such as baryon number, lepton number, or shape multipoles --- is radiated away during gravitational collapse~\cite{Bekenstein:1998aw}. 

This theorem simplifies the study of PBH evolution. Regardless of the complex particle physics interactions during the early Universe, the resulting \ac{BH} is fully described by $(M, J, Q)$. For astrophysical purposes, $Q$ is typically negligible due to rapid charge neutralization by the surrounding plasma, leaving $M$ and $J$ as the primary parameters. Furthermore, this uniqueness is the foundation for \ac{BH} Thermodynamics~\cite{1973-Bekenstein-PRD}, where the horizon area corresponds to entropy, a crucial concept for understanding PBH evaporation. However, note that \ac{BH} solutions of non-linear matter or higher-order gravity corrections evade the classical no-hair theorems~\cite{Herdeiro:2015waa,Antoniou:2017acq,Doneva:2017bvd,Silva:2017uqg,Bakopoulos:2023fmv,Parvez:2025wtq}.

If the scalar fields during inflation couple non-minimally to gravity or possess specific self-interactions, they can evade the classical \emph{No-Hair Theorems}. This leads to the phenomenon of spontaneous scalarization or the formation of \emph{scalar hair} around the \ac{BH}~\cite{Sotiriou:2014pfa, Herdeiro:2015waa}. This is phenomenologically highly significant: if \acp{PBH} inherit scalar hair from the inflationary epoch, their physical properties will fundamentally differ from late-time astrophysical black holes. Specifically, the presence of scalar hair modifies the spacetime geometry near the event horizon, which can significantly alter the \ac{PBH} accretion rates in the early universe and lead to distinct deviations in their Hawking evaporation signatures~\cite{Kanti:2004nr}. Consequently, searching for these anomalous evaporation spectra or accretion-driven feedback could provide a unique observational probe to distinguish genuine \acp{PBH} from stellar-collapse \acp{BH}, while simultaneously constraining modified gravity theories.

Note that when applying the classical \emph{No-Hair Theorem} to \acp{PBH}, the extreme conditions of the early universe impose strict physical constraints that effectively reduce this parameter space. First, the primordial plasma in the \ac{RD} era was highly coupled and possessed an immense electrical conductivity. Any transient charge fluctuation acquired during the collapse would be instantaneously neutralized by the rapid accretion of oppositely charged particles from the surrounding thermal bath \cite{DeLuca:2020fpg}. Consequently, the electric charge of \acp{PBH} is universally expected to be negligible ($Q \approx 0$).

Second, the generation of angular momentum during PBH formation is highly restricted. In the standard scenario of collapse from inflationary fluctuations, the contracting Hubble patch is highly spherically symmetric. At linear (first) order in curvature perturbations, there is no mechanism to break this symmetry and provide a tidal torque; therefore, PBHs are theoretically predicted to be born with near-zero initial spin ($J \approx 0$). However, rigorous modern treatments reveal that this is not strictly zero. When second-order cosmological perturbations and potential non-Gaussianities are incorporated, subtle tidal torques from neighboring overdensities impart a small but non-vanishing initial spin, typically peaking at a dimensionless spin parameter $a^* \sim \mathcal{O}(0.01)$ \cite{Mirbabayi:2019uph, DeLuca:2019buf}.

This nuanced understanding of initial PBH spin is of paramount phenomenological importance. In recent years, \ac{GW} detections by the \ac{LVK} collaboration have revealed a population of massive binary black hole mergers exhibiting surprisingly low effective inspiral spins ($\chi_{\text{eff}} \approx 0$). While standard astrophysical binary evolution models often struggle to naturally reproduce such persistently low spins without fine-tuning, it is an automatic, fundamental prediction of the PBH scenario~\cite{Garcia-Bellido:2017mdw, Fernandez:2019kyb}. Thus, the near-zero birth spin dictated by cosmological perturbation theory serves as one of the most compelling observational signatures supporting a primordial origin for a subset of the LVK events.

\section{Cosmological Black Holes}
\label{sec:CosmoBH}

The \acp{BH} discussed in Sec.~(\ref{sec:SphSy}) (Schwarzschild/Kerr) share a fundamental simplification: they are \emph{isolated} in an asymptotically flat universe. However, \acp{PBH} form and evolve within the dense, rapidly expanding plasma of the early Universe, described by the \ac{FLRW} metric. Thus, the 
line-element in Eq. \eqref{eq:gen-4D} should be generalized in order to describe a \ac{PBH}: 
\begin{eqnarray}
\label{eq:gen-4D-TD}
ds^2 &=& - f(r,t) dt^2 + \frac{dr^2}{g(r,t)} + R^2(r,t) d\Omega^2 \, .
\end{eqnarray}
where $f(r,t)$, $g(r,t)$ and $R(r,t)$ (areal radius) are arbitrary smooth functions of the radial coordinate $r$ and $t$.

This non-isolated environment introduces fundamental challenges to the standard definitions:
\begin{enumerate}
    \item \textbf{Dynamic Background:} The spacetime is not static ($R_{\mu\nu} \neq 0$), so the Time-like Killing vector $\xi^{\mu}$ does not strictly exist.
    \item \textbf{Accretion:} PBHs are not in vacuum; they interact with the surrounding radiation fluid, making the mass $M(t)$ time-dependent.
    \item \textbf{Asymptotics:} The metric does not approach Minkowski space at infinity; it approaches the FLRW background.
\end{enumerate}

Consequently, theorems relying on stationarity --- such as the standard No-Hair theorem --- do not strictly apply during the dynamic formation and accretion phases~\cite{1969-Lynden_Bell-Nature,Herdeiro:2015waa}. To describe these objects, several exact solutions embedding \acp{BH} in cosmological backgrounds have been proposed, such as the McVittie solution~\cite{1933-McVittie-MNRAS}, the Sultana-Dyer solution~\cite{2005-sultana_Dyer-GRG,Xavier:2021chn}, and Lemaître-Tolman-Bondi (LTB) models~\cite{Faraoni:2021nhi}.

However, rather than relying on specific exact solutions, the modern study of PBHs utilizes two robust quasi-local definitions to characterize horizons and mass in time-dependent spacetimes.

\subsection{The Apparent Horizon}
In a dynamic universe, the classic \emph{Event Horizon} is teleological --- it requires knowledge of the entire future history of the spacetime to define. This is impractical for PBH formation. Instead, we use the \emph{Apparent Horizon}, which is a local concept~\cite{Ashtekar:2025wnu}.

The apparent horizon is defined as the outermost \ac{MTS}. Mathematically, for a spherically symmetric metric with areal radius $R(t,r)$, the apparent horizon is located where the expansion of the outgoing null geodesics ($\theta_{out}$) vanishes:
\begin{equation}
g^{\mu\nu} \nabla_\mu R \nabla_\nu R = 0 \quad \implies \quad \nabla_\mu R \text{ is null.}
\end{equation}
In practical terms, for a collapsing region in an expanding universe, the apparent horizon marks the boundary where gravity becomes strong enough to stop light from moving outward relative to the local expansion \cite{Faraoni:2015ula}. Unlike the event horizon, the apparent horizon can evolve discontinuously during violent formation processes~\cite{Booth:2005qc,Ashtekar:2025wnu}.

\subsection{The Misner-Sharp Mass and Kodama Vector}
\label{sec:MSDef-Kodama}

In stationary asymptotically flat spacetimes, the total energy (Komar mass) is rigorously defined by integrating the conserved current associated with the time-like Killing vector.   In dynamic, spherically symmetric cosmological spacetimes (where no time-like Killing vector exists), the Kodama vector $K^\mu$ serves as its unique dynamic generalization~\cite{1980-Kodama-PTP}.

Since we cannot define a conserved \emph{Schwarzschild Mass} in the standard way, the appropriate analog for spherically symmetric, dynamic spacetimes is the \emph{Misner-Sharp Mass} ($M_{MS}(r,t)$) \cite{1964-Misner_Sharp-PRD}. It is defined purely geometrically by the curvature of the 2-sphere:
\begin{equation}
M_{MS}(r,t) = \frac{R}{2G} \left( 1 - g^{\mu\nu} \nabla_\mu R \nabla_\nu R \right) \, .
\end{equation}
Physically, $M_{MS}$ represents the total energy (matter + gravitational) enclosed within the sphere of radius $R$.
This mass is associated with the \emph{Kodama Vector} $K^\mu$, which serves as a dynamic replacement for the Killing vector. $K^\mu$ is defined as~\cite{1980-Kodama-PTP}:
\begin{equation}
K^\mu = \epsilon^{\mu\nu} \nabla_\nu R \, ,
\end{equation}
where $\epsilon^{\mu\nu}$ is the volume form of the 2-dimensional quotient space perpendicular to the 2-spheres. Using the Einstein field equations, we construct the Kodama energy current:
\begin{equation}
    J^{\mu} = G^{\mu\nu}K_{\nu} = 8\pi G \, T^{\mu\nu}K_{\nu} \, .
\end{equation}
As a direct consequence of the Bianchi identity ($\nabla_{\mu}G^{\mu\nu}=0$) and the symmetries of the spacetime, the Kodama current is covariantly conserved:
\begin{equation}
    \nabla_{\mu}J^{\mu} = 0 \, .
\end{equation}
Because the current is divergenceless, we can use Gauss's theorem to define a conserved geometric charge $Q$ within a spatial volume $\Sigma$ bounded by a 2-sphere $\partial \Sigma$ of areal radius $R$:
\begin{equation}
    Q = \int_{\Sigma} J^{\mu} d\Sigma_{\mu} \, .
\end{equation}
By explicitly integrating this over the spherically symmetric metric, one finds that the conserved charge is identically proportional to the Misner-Sharp mass:
\begin{equation}
    Q = 8\pi G M_{MS}(R) \, .
\end{equation}
This mathematical identity guarantees that the Misner-Sharp mass is not just a convenient geometric scalar, but the actual conserved Noether-like charge of the dynamic spacetime. Therefore, when $M_{MS}$ is evaluated at the apparent horizon during PBH formation, the resulting mass measurement is completely gauge-invariant and entirely free from coordinate artifacts~\cite{1980-Kodama-PTP,2006-Racz-CQG,Abreu:2010ru}.

\subsection{Conceptual Pitfalls: Horizons and Mass Definitions}
\label{sec:Pitfalls}

Transitioning from vacuum gravity to cosmological collapse introduces two common sources of confusion regarding the behavior of horizons and the rigorous definition of mass.

\subsubsection{The Tale of Three Horizons: Re-entry, Sound, and Formation}

A frequent point of confusion in the PBH literature arises from the generic term \emph{horizon crossing.} In standard cosmology, the Hubble Horizon (or cosmological horizon), $R_H \sim H^{-1}$, expands linearly with time during the radiation and matter eras ($R_H \propto t$). Modes (perturbations) that exited the horizon during inflation subsequently \emph{re-enter} the horizon as the universe expands.

One might ask: \emph{If the cosmological horizon is expanding outward, how does a \ac{BH} horizon form by matter collapsing inward, and what prevents it from simply bouncing back?}

The resolution lies in carefully distinguishing the interplay between \emph{three} distinct physical horizons:
\begin{enumerate}
\item \textbf{Cosmological Horizon ($R_H$):} This represents the absolute causal sphere of influence. It strictly expands ($dR_H/dt > 0$). When a primordial perturbation of scale $\lambda$ re-enters ($ \lambda \lesssim R_H$), gravity can finally act causally across the entire density fluctuation to initiate collapse.
\item \textbf{Sound Horizon ($R_s$):} This is the distance a pressure wave can travel since the Big Bang, defined approximately as $R_s \approx c_s R_H$. In the \ac{RD} epoch, the sound speed is extremely high ($c_s = 1/\sqrt{3}$), making the sound horizon only slightly smaller than the cosmological horizon. The sound horizon dictates the causal reach of \emph{pressure gradients}. If gravity does not trigger collapse fast enough upon re-entry, the sound horizon overtakes the perturbation, allowing pressure forces to violently push the overdensity apart into acoustic oscillations.
\item \textbf{\ac{BH} (Apparent) Horizon ($R_{AH}$):} If the fluctuation amplitude $\delta$ is large enough upon re-entry, local gravity overwhelms the pressure forces dictated by the sound horizon. While the background Hubble horizon continues to expand outward, the local spacetime region decouples from the Hubble flow and collapses inward. A \emph{new} trapped surface forms locally, representing the apparent horizon of the nascent black hole.
\end{enumerate}
At this juncture, it is physically instructive to distinguish between the topological classifications of the resulting \acp{BH}, which depend sensitively on the initial perturbation amplitude. The standard collapse scenario described 
above --- where the physical areal radius of the spacetime monotonically increases with the comoving radial coordinate --- is classified as a \emph{Type I} fluctuation. This leads to standard trapping horizons and represents the standard phenomenological focus of most PBH literature.

However, for sufficiently large initial amplitudes, the local spatial geometry becomes highly positively curved, developing a \emph{throat or neck} structure where the areal radius behaves non-monotonically. These are classified as \emph{Type II} fluctuations. While historically treated as mathematically pathological or exceptionally rare, this topological complexity becomes highly relevant in non-Gaussian inflationary scenarios where the primordial Probability Density Function exhibits heavy tails, naturally producing these extreme amplitudes~\cite{Ferrante:2022mui}. 

As recently demonstrated by high-resolution \ac{NR} simulations~\cite{Uehara:2024yyp}, the gravitational collapse of these Type II perturbations exhibits a distinct configuration characterized by bifurcating trapping horizons (also referred to as Type B PBHs). Because standard apparent horizon finders can fail or misidentify these complex, baby-universe-like topologies, analyzing Type II PBHs requires highly sophisticated \ac{NR} tracking methods. Nonetheless, because the vast majority of standard inflationary models heavily suppress these extreme amplitudes, the bulk of this review will focus on the formation mechanics of Type I PBHs.

PBH formation is therefore a deeply relativistic hydrodynamic phenomenon: it is the creation of a \emph{small, local} apparent horizon, driven by gravity winning the "race" against the sound horizon, all nested deep \emph{within} the much larger, expanding cosmological horizon.

\subsubsection{The Problem of Mass in a Non-Empty Universe}

In the Schwarzschild solution, mass ($M$) is an integration constant defined at spatial infinity where the energy density vanishes ($T_{\mu\nu} \to 0$). In the early Universe, this definition fails because the background density is non-zero everywhere. If one were to integrate the density to infinity, the mass would diverge:
\begin{equation}
M(r) \sim \int_0^r 4\pi \rho_{\text{bg}} \, r'^2 dr' \propto r^3 \to \infty \, .
\end{equation}
\textit{Question: How do we distinguish the \emph{\ac{BH} Mass} from the \emph{Universe's Mass}?}

The rigorous solution utilizes the Misner-Sharp Mass $(M_{MS})$ evaluated specifically at the Apparent Horizon $(R_{AH})$. At the location of the horizon, the compactness condition is exactly satisfied:
\begin{equation}
    2GM_{MS}(R_{AH})=R_{AH} \, .
\end{equation}
Crucially, the physical validity of this measurement is underpinned by the Kodama vector introduced in Section \eqref{sec:MSDef-Kodama}. In a non-stationary cosmological background, attempting to separate the mass of a localized BH from the expanding cosmic fluid is notoriously prone to coordinate gauge ambiguities. However, because the Misner-Sharp mass is mathematically equivalent to the conserved charge associated with the Kodama current ($J^{\mu}=G^{\mu\nu}K_{\nu}$), evaluating it at the apparent horizon ensures that the measured energy is not a mere artifact of the coordinate choice. It is a strictly gauge-invariant quantity representing the true total mass (matter plus gravitational field) enclosed within the trapped surface (see Sec.~\eqref{sec:MSDef-Kodama} for the mathematical derivation).

However, a subtlety remains: does this mass include the background radiation? Yes. This leads to two practical ways of tracking the mass:
\begin{description}
    \item[\textbf{Mass Excess}] Some literature defines the PBH mass via the mass excess $\delta M(r,t) = M_{MS}(r,t) - M_{bg}(r,t)$, subtracting the background FLRW mass that \emph{would have been there} in the absence of a perturbation. Because $M_{MS}$ is tied to the physically conserved Kodama flow, the mass excess $\delta M$ rigorously and gauge-invariantly quantifies the true localized energy responsible for the collapse. As we will show in Sec.~\eqref{sec:Compaction}, the gauge-invariance is the key factor in 
    using compaction function $\mathcal{C}$ as a measure for the threshold for PBH formation.
    \item[\textbf{Operational Definition}] Once the initial collapse is complete, the resulting black hole effectively decouples from the background Hubble flow. While a strict vacuum does not form --- as the universe remains permeated by the continuous cosmic fluid --- the rapid localized infall of material creates a severe underdense \emph{depletion zone} (or rarefaction wave) immediately outside the apparent horizon. Because the black hole is now dynamically detached from the global cosmic expansion, the apparent horizon rapidly stabilizes. For practical and computational purposes, the Misner-Sharp mass evaluated at this stabilized boundary is operationally defined as the intrinsic mass of the \ac{PBH}, effectively isolating the remnant from the surrounding cosmological environment. We will discuss this more in Sec. \eqref{sec:SHorizon-MSharpMass}.
\end{description}

\section{The Mechanics of Collapse}
\label{sec:FormationMechanics}

\textit{If the \ac{RD} early universe was so dense, why didn't the whole thing collapse into a \ac{BH} immediately?}

To fully resolve this apparent conundrum, we must first understand the evolution of the cosmological background --- specifically the causal horizon and the pressure forces at play --- before examining the behavior of local density perturbations.
As mentioned earlier, in this section, we explicitly include speed of light ($c$).

\subsection{The Mass-Time Relation}
\label{sec:mass_time}
In the \ac{RD} era, the scale factor evolves as $a(t) \propto t^{1/2}$, and the Hubble parameter is $H = {1}/{(2t)}$. The Hubble Radius (the causal horizon) is given by:
\begin{equation}
R_H \approx \frac{c}{H} = 2ct \, .
\end{equation}
The total mass enclosed within this causal sphere, $M_H$, is the physical volume multiplied by the \emph{threshold mass density} $\rho_c = {3H^2}/{(8\pi G)}$:
\begin{equation}
M_H = \frac{4\pi}{3} \left( \frac{c}{H} \right)^3 \left( \frac{3H^2}{8\pi G} \right) = \frac{c^3}{2GH} \, .
\end{equation}
Substituting $H = 1/(2t)$, we obtain a simple linear relation between the horizon mass $(M_H)$ and cosmic time:
\begin{equation}
M_H(t) =  \frac{c^3}{G} t \, .
\end{equation}
Expressing this in astrophysical units ($M_{\odot}$ and seconds), we find:
\begin{equation}
\label{eq:MassTime}
M_H(t) \approx 2 \times 10^5 M_{\odot} \left( \frac{t}{1 \text{ s}} \right) \, .
\end{equation}
This linear scaling implies a strict \textbf{Mass-Time Lock}:
\begin{itemize}
    \item At $t \sim 10^{-15}$ s, $M_H \sim 10^{-10} M_{\odot}$ (Asteroid mass).
    \item At $t \sim 10^{-5}$ s, $M_H \sim 1 M_{\odot}$ (Stellar mass).
    \item At $t \sim 1$ s, $M_H \sim 10^5 M_{\odot}$ (Supermassive).
\end{itemize}

\subsection{The initial minimum mass formation constraint}

An immediate physical consequence of this strict temporal lock is the existence of a fundamental lower bound on the PBH mass spectrum at the time of formation. Because the exponential expansion during inflation radically dilutes any pre-existing topological defects, relics, or BHs, the very first phenomenologically relevant PBHs can only form immediately after inflation ends, precisely as the universe enters the \ac{RD} era. 

Consequently, the absolute minimum possible \emph{initial formation mass} for a standard PBH is dictated by the horizon mass at the reheating epoch~\cite{Bassett:2005xm,Ozsoy:2023ryl,Lozanov:2019jxc}. For a standard high-scale inflationary model with a reheating temperature of $T_{\text{RH}} \sim 10^{15}$ GeV, this initial horizon mass evaluates to approximately $M_{\text{min}} \sim 1$ g \cite{Carr:2020xqk}. If the early universe underwent a phase of low-scale inflation~\cite{German:2001tz,Mathew:2016anx}, or if the reheating temperature was significantly lower, this minimum mass cutoff would be correspondingly higher. Any BH discovered with a mass smaller than this inflationary threshold would require either highly exotic post-inflationary physics or a fundamentally different formation mechanism.

However, one must carefully distinguish between the mass of a PBH at formation and its mass observed today. Because PBHs undergo Hawking evaporation (we will discuss in Sec. \eqref{sec:HawkingRad}), a BH that initially formed with a much larger mass (e.g., $\sim 10^{15}$ g) will continually loose mass over cosmic time. Furthermore, if the late stages of evaporation are drastically slowed due to the FLRW background~\cite{Xavier:2021chn} or halted by quantum backreaction~\cite{Mandal:2023kpu} --- such as the memory burden effect discussed previously --- these BHs can survive as tiny, sub-gram remnants. Therefore, the discovery of a microscopic BH today would not necessarily violate the inflationary horizon bound. Instead, it would imply that the object is either an evaporated remnant of a heavier PBH, or the product of a fundamentally different, non-cosmological formation mechanism.

While analytical frameworks such as the Press-Schechter formalism and peak theory provide critical qualitative intuition, they fundamentally struggle to capture the complex, highly non-linear gravitational dynamics that govern the final stages of primordial collapse. To circumvent these analytical limitations, significant effort has recently been funneled into high-fidelity numerical relativity simulations. These numerical approaches solve the full, un-approximated Einstein field equations in expanding spacetimes to precisely chart the transition from super-horizon primordial fluctuations to fully formed black hole horizons. A prominent example of this push is the development of specialized, open-source computational toolkits like \texttt{COSMOS}~\cite{Yoo:2026itl}, \texttt{SpriBHoS}~\cite{Escriva:2019nsa, Escriva:2025rja} (see also Ref.~\cite{Musco:2018rwt}), a $3+1$-dimensional numerical relativity package engineered specifically to handle the non-linearities of PBH formation using advanced features like non-Cartesian scale-up coordinates and localized mesh refinements. These ongoing numerical endeavors continue to refine our tracking of the critical threshold $\delta_c$ and the profile-dependent critical scaling laws, bridging the gap between primordial initial conditions and exact horizon metrics. To get some physical insight about the PBH formation and the physics behind, this review focuses on semi-analytical frameworks.

\subsection{The Jeans Barrier and the Race Against Sound}
\label{sec:JeansBarrier}

One might ask: \emph{Why can not we form a small \ac{BH} of mass $10^{-10} M_{\odot}$ at $t=1$ second?} The causal horizon at $t=1$ s is certainly large enough ($10^5 M_{\odot}$) to contain such a tiny mass.

The answer lies in the dynamics of the \textbf{Jeans Length} ($\lambda_J$) and a fundamental physical mechanism: the \emph{race against sound}. As illustrated in Fig.~\eqref{fig:PBHformation}, in the \ac{RD} era, the primordial plasma is highly relativistic, and pressure forces are enormous ($P = \rho c^2 / 3$). For details, see Appendix~\eqref{app:JeansDetail}. For local self-gravity to trigger a collapse, the physical size of the density perturbation ($\lambda$) must exceed the physical Jeans length~\cite{Niemeyer:1999ak}:
\begin{equation}
\lambda_J^{\text{phys}} = c_s \sqrt{\frac{\pi}{G \rho}} \, ,
\end{equation}
where $c_s \approx c/\sqrt{3}$ is the sound speed. Because the background mass density scales as $\rho \propto a^{-4}$, the physical Jeans length scales as $\lambda_J \propto a^2$. Similarly, the physical Hubble horizon scales as $R_H \propto H^{-1} \propto a^2$. 

Comparing the two reveals a tight, epoch-independent constraint. In a pure radiation fluid, the Jeans length is always a fixed, large fraction of the Hubble horizon:
\begin{equation}
\lambda_J^{\text{phys}} = \frac{1}{\sqrt{3}} R_H \simeq 0.58 R_H \, .
\end{equation}
To understand why this locks the PBH mass to the formation time and dictates the strict threshold for collapse, we must track the lifecycle of a comoving density perturbation \cite{Carr:1975qj, Sasaki:2018dmp}:
\begin{description}
    \item[\textbf{Super-horizon phase ($\lambda > R_H$)}] Early on, the physical wavelength of the perturbation is larger than the causal horizon. The region is causally disconnected, preventing any fluid dynamics (like collapse or sound waves) from acting. The perturbation is essentially "frozen" into the spacetime metric.

\item[\textbf{Horizon Re-entry ($\lambda \sim R_H$)}] 
As illustrated in the top-right in Fig.~\eqref{fig:PBHformation}, while the perturbation enters the horizon, it becomes causally connected. At this precise moment, a cosmic race begins between gravity (attempting to collapse the region in a free-fall time, $t_{\text{ff}} \sim 1/\sqrt{G\rho}$) and pressure (attempting to disperse the region via sound waves in a sound-crossing time, $t_s \sim \lambda / c_s$). 
    
As derived in Appendix \eqref{app:Jeans-RaceAgainstTime}, a region only collapses if its gravity is strong enough to crush the sphere (time of collapse or $t_{\rm coll}$)  before internal pressure waves can propagate to the boundary ($t_{\rm cross} \approx R/c$~s) to disperse the overdensity. Since sound in a radiation fluid travels at $c_s \approx  c/\sqrt{3}$, only the most extreme overdensities ($\delta >1/3$) possess the gravitational acceleration necessary to win this race. 
    
Because the primordial fluid is \ac{RD}, the pressure waves travel at relativistic speeds ($c_s \approx 0.58 c$). For the overdensity to successfully collapse before a sound wave can travel from its center to its edge and blow it apart, the local gravity must be overwhelmingly strong. This physical requirement—that the free-fall time must be shorter than the sound-crossing time ($t_{\text{ff}} \lesssim t_s$)—naturally establishes the critical density threshold for PBH formation. In a generic fluid, this yields $\delta_c \sim c_s^2 / c^2 \sim w$. As illustrated in bottom-right Fig.~\eqref{fig:PBHformation}, only the most extreme, highly compact fluctuations possess enough self-gravity to overcome these nearly speed-of-light pressure gradients. This incredibly brief temporal window $(0<t_{\rm ff}<t_s)$, combined with the relativistic sound speed, is precisely why PBH formation is such an exceedingly rare cosmological event.
    
\item[\textbf{Sub-horizon phase ($\lambda \ll R_H$)}] If the perturbation fails to collapse immediately upon re-entry, the expanding universe rapidly changes the balance. As the horizon and Jeans length continue to grow as $a^2$, they quickly outpace the perturbation size ($\lambda \propto a$). The perturbation soon becomes much smaller than the Jeans length ($\lambda \ll \lambda_J$). At this point, the relativistic pressure gradients utterly dominate gravity, and the overdensity is permanently dispersed into acoustic oscillations \cite{Sasaki:2018dmp}.
\end{description}

In conclusion, a PBH cannot form from a sub-horizon fluctuation because the Jeans barrier absolutely dominates on small scales. Returning to our initial question: A $10^{-10} M_{\odot}$ patch of space crossing the horizon at $t \sim 10^{-15}$ s has a fleeting chance to collapse if it can win the race against relativistic sound. If it fails, it turns into sound waves. By $t=1$ s, the physical size of that $10^{-10} M_{\odot}$ region is microscopically small compared to the Jeans length at that epoch ($\lambda \ll \lambda_J$). The pressure barrier makes it entirely impossible for such a small mass to spontaneously collapse at a later time. Therefore, the mass of a PBH is inextricably locked to its exact time of horizon re-entry ($M_{\text{PBH}} \sim M_H(t_{\rm cross})$).

\begin{table*}[!htb]
\centering
\renewcommand{\arraystretch}{1.2}
\begin{tabular}{|c|c|c|c|c|}
\hline
\textbf{Time} ($t$) & \textbf{Mass} ($M_H$) & \textbf{BH Density} ($\epsilon_{\text{BH}}$) & \textbf{Univ. Density} ($\rho_{\rm rad}$) & \textbf{Ratio} $= \frac{\epsilon_{\text{BH}}}{\rho_{\text{rad}}}$ \\
\hline
$10^{-15}$ s & $10^{-10} M_{\odot}$ & $10^{25}$ & $10^{23}$ & $\approx 56$ \\
$10^{-5}$ s & $1 M_{\odot}$ & $10^{5}$ & $10^{3}$ & $\approx 56$ \\
$1.0$ s & $10^{5} M_{\odot}$ & $10^{-5}$ & $10^{-7}$ & $\approx 56$ \\
\hline
\end{tabular}
\caption{Comparison of the absolute density of the formed \ac{BH} ($\epsilon_{\text{BH}}$) and the background radiation density ($\rho_{\rm rad}$) in $\mathrm{MeV}/\mathrm{fm}^3$. The ratio implies an efficiency factor of $\gamma \approx 0.13$.}
\label{tab:pbh_density_mev}
\end{table*}

\subsection{Scale Invariance and the Kinematic Threshold}
\label{sec:scale_invariance}

Having established that the mass of a PBH scales with the horizon mass, we must refine this relationship: a collapsing region does not swallow the entire causal horizon. Instead, the final mass of the \ac{BH} is a fraction of the horizon mass at the time of re-entry:
\begin{equation}
    M_{\text{PBH}} = \gamma M_H(t) \, ,
\end{equation}
where $\gamma < 1$ is the collapse efficiency factor, which depends on the fluid dynamics and the equation of state of the background fluid~\cite{Carr:1975qj,Niemeyer:1999ak,Byrnes:2018clq}.

Using Eq.~\eqref{eq:MassTime}, we can map the background radiation energy density $\rho_{\text{rad}}$ to the horizon mass. 
\begin{equation}
\rho_{\text{rad}} \simeq 10^4 \left(\frac{M_H}{M_{\odot}}\right)^{-2} \quad \frac{\mathrm{MeV}}{\mathrm{fm}^3} \, .
\end{equation}
where $\mathrm{fm}$ is femto-meter. Table (\ref{tab:pbh_density_mev}) compares the absolute energy density of the background universe ($\rho_{\rm rad} \sim H^2$) to the absolute internal density of the resulting \ac{BH} ($\epsilon_{\text{BH}} \sim M_{\text{PBH}}/R_S^3$) at various epochs. 

Let us try to physically interpret the ratio ($\epsilon_{\text{BH}}/\rho_{\text{rad}}$) in Table \eqref{tab:pbh_density_mev}.  Because the density of a \ac{BH} scales inversely with the square of its mass ($\epsilon_{\text{BH}} \propto M^{-2}$), the ratio of the \ac{BH} density to the background density is purely a function of the efficiency factor:
\begin{equation}
    \frac{\epsilon_{\text{BH}}}{\rho_{\text{rad}}} = \left( \frac{M_H}{M_{\text{PBH}}} \right)^2 = \frac{1}{\gamma^2} \, .
\end{equation}
The constant ratio of $\approx 56$ in Table (\ref{tab:pbh_density_mev}) corresponds to an efficiency factor of $\gamma = 1/\sqrt{56} \approx 0.13$. This physically signifies that roughly $13\%$ of the mass within the Hubble patch \emph{can at-max} successfully collapse to a BH, while the rest is blown away by the intense radiation pressure gradients during formation \cite{Niemeyer:1999ak}.

Furthermore, the fact that this ratio remains strictly constant demonstrates that the \emph{relative kinematic threshold} for PBH formation is perfectly scale-invariant during a pure \ac{RD} epoch ($a \propto t^{1/2}$). Because both densities scale exactly as $t^{-2}$, the necessary fractional overdensity ($\delta_c \sim \Delta \rho / \rho$) to trigger the collapse is identical across all mass scales. 

However, the kinematic threshold alone cannot determine the final mass spectrum of PBHs. The actual abundance at any given scale depends heavily on two factors --- the primordial power spectrum and equation of state variation. 
The statistical probability that a local region possesses the required overdensity ($\delta > \delta_c$) is dictated by the inflationary dynamics. We will discuss more on this in Sec. \eqref{sec:Compaction}.

While the preceding discussion assumes a perfectly scale-invariant, \ac{RD} background with a constant equation of state ($w \approx 1/3$) and a constant sound speed ($c_s \approx c/\sqrt{3}$), the true thermal history of the early universe is far more dynamic. As the universe expands and cools, the primordial plasma undergoes several major thermal phase transitions as different particle species become non-relativistic and annihilate.

The most phenomenologically significant of these events is the Quantum Chromodynamics (QCD) phase transition, occurring at a temperature of $T \sim \mathcal{O}(100)$ MeV. During this epoch, free quarks and gluons become confined into hadrons, causing a rapid drop in the effective number of relativistic degrees of freedom ($g_*$). This thermodynamic shift causes a transient ``softening" of the plasma, leading to a localized dip of approximately $10\% - 30\%$ in both the equation of state parameter $w$ and the sound speed squared $c_s^2$ \cite{Byrnes:2018clq}.

From the discussion in the previous subsection \eqref{sec:JeansBarrier} race-against-sound derivation that the critical density threshold for collapse is strictly determined by the sound speed ($\delta_c \sim (c_s/c)^2$). Therefore, this transient softening of the plasma fundamentally alters the mechanics of collapse. The drop in pressure support effectively lowers the Jeans barrier, making it significantly easier for overdensities to collapse against the weakened outward pressure gradients. 

Because the probability of PBH formation is highly sensitive to the critical threshold, even a modest reduction in $\delta_c$ results in a massive amplification of the formation fraction $\gamma$. Consequently, the efficiency $\gamma$ presented in standard scale-invariant models (such as in Table \eqref{tab:pbh_density_mev}) is not strictly a constant, but rather becomes a time-dependent function, $\gamma(t)$, strongly peaking during thermal transitions.

Crucially, the horizon mass at the precise time of the QCD transition evaluates to approximately $1 - 2 \, M_{\odot}$~\cite{Jedamzik:1996mr, Byrnes:2018clq}. As a direct consequence of this thermodynamic softening, the universe naturally imprints a prominent "bump" in the PBH mass function at the solar mass scale~\cite{Jedamzik:1996mr, Byrnes:2018clq}. This theoretically preferred mass scale provides a compelling natural explanation for the abundance of compact objects detected by the LVK collaboration, bridging the mechanics of the QCD vacuum with GW astronomy~\cite{Cang:2023ysz}.

\subsection{Primordial Spin Statistics}
\label{sec:PBHSpin}

A fundamental property distinguishing \acp{PBH} from their astrophysical counterparts is the distribution of their dimensionless spin parameter, $a_* \equiv J/(GM^2)$. While astrophysical BHs formed from the collapse of rotating stellar cores obey the conservation of angular momentum and typically retain high spins ($a_* \sim 0.7 - 0.9$), PBHs formed in the standard \ac{RD} scenario are theoretically predicted to be born with negligible rotation~\cite{DeLuca:2019buf,Mirbabayi:2019uph}.

\subsubsection{The Zero-Spin Approximation}

In the standard formation scenario, PBHs form from the direct collapse of curvature perturbations $\zeta$ upon horizon entry. At zeroth order in perturbation theory, the collapsing Hubble patch is modeled as a spherically symmetric peak in a Gaussian random field. Due to the isotropy of the background radiation fluid, there is no physical mechanism to generate a preferred axis of rotation or provide a tidal torque. Consequently, the resulting BH is Schwarzschild (non-rotating) to leading order~\cite{Chiba:2017rvs}.

\subsubsection{First-Order Corrections: Tidal Torques}

However, perfectly spherical collapse is a mathematical idealization. In reality, PBH formation sites correspond to local maxima (or \emph{peaks}) within the stochastic primordial density field~\cite{DeLuca:2019buf,Mirbabayi:2019uph}. Rigorous statistical treatments of these peaks demonstrate that the initial spin of the resulting \ac{BH} is not strictly zero. Non-vanishing angular momentum is generated at first order by tidal torques --- specifically, the classical coupling between the inertia tensor of the slightly asymmetric collapsing patch and the tidal gravitational fields induced by neighboring density fluctuations. Analytical calculations by De Luca \textit{et al.}~\cite{DeLuca:2019buf} and Mirbabayi \textit{et al.}~\cite{Mirbabayi:2019uph} quantify this effect. For a standard \ac{RD} equation of state ($w=1/3$), the dimensionless spin ($a_*$) follows a Rayleigh distribution characterized by the root-mean-square expectation value:
\begin{equation}
    \sqrt{\langle a_*^2 \rangle} \approx \frac{1}{\gamma} \sqrt{ \Omega_{\text{rad}} } \sigma_{\delta} \sim \mathcal{O}(10^{-2}) \, .
\end{equation}
Here, $\gamma$ is the collapse efficiency factor, $\Omega_{\text{rad}} \approx 1$ is the background energy density parameter for radiation, and most importantly, $\sigma_{\delta}$ is the root-mean-square amplitude (variance) of the density contrast evaluated at the horizon scale. Because PBHs form from exceptionally rare, high-amplitude fluctuations ($\delta \sim \delta_c \gg \sigma_{\delta}$), the background variance $\sigma_{\delta}$ must naturally be small (typically $\sim 10^{-2}$) to avoid overproducing them and violating observational bounds. Because the angular momentum scales linearly with this variance, the tidally induced spin is heavily suppressed. Thus, standard PBHs are born effectively non-spinning ($a_* < 0.01$). This prediction is robust against changes in the power spectrum shape but relies heavily on the baseline assumption of Gaussianity.

\subsubsection{The Matter-Domination Exception}

A critical caveat exists for PBHs formed during an \ac{EMD} era --- for example, during a prolonged reheating phase or a phase transition. In an EMD ($w=0$), the pressure gradients that usually resist collapse are absent. Consequently, the collapse is highly non-spherical, and the breakdown of the Jeans criterion allows anisotropy to grow non-linearly. Harada et al.~\cite{Harada:2017fjm} demonstrated that PBHs formed in these epochs can possess significant initial spins, with $a_*$ potentially approaching unity.

\subsubsection{Observational Discriminator}
The \emph{near-zero} birth spin of \ac{RD} era PBHs serves as a potent multimessenger discriminator. 
\begin{description}
\item[\textbf{Astrophysical BHs}] Stellar evolution models and binary synthesis codes predict a broad spin distribution, often peaking at $a_* \approx 0.7$~\cite{Sasaki:2018dmp}.
\item[\textbf{Primordial BHs}] Predict a narrow distribution effectively at $a_* \approx 0$.
\end{description}
As mentioned in the introduction, recent observations by the LVK collaboration of binary black hole mergers with low effective inspiral spins ($\chi_{\text{eff}} \approx 0$) align with the primordial prediction. While late-time accretion can theoretically spin up a PBH (creating a thin disk that transfers angular momentum), this process is generally inefficient for PBHs that remain in the diffuse halo environment, preserving their "fossil" low-spin status~\cite{Fernandez:2019kyb}.

\subsection{Absolute Density vs. Density Contrast: Resolving the Conundrum}
\label{sec:density_contrast}

We now return to the pedagogical question posed at the beginning of this section: \emph{Why did the entire cosmos not immediately collapse into a \ac{BH}, given that its absolute average density in the early universe vastly exceeded the density of a typical \ac{BH} today?}

The resolution lies in understanding the competing physical forces: local self-gravity versus radiation pressure and the cosmological expansion. As established above, during the \ac{RD} era, both the background energy density and the radiation pressure ($P = \rho/3$) scale as $t^{-2}$. As we look further backward in time, the universe was characterized by enormously high absolute densities and correspondingly intense outward radiation pressure. 

For a local region to decouple from the background and collapse into a PBH, its local self-gravity must be strong enough to overcome both the outward radiation pressure and the rapid background Hubble expansion ($H \propto t^{-1}$). Because the radiation pressure increases so dramatically at earlier times, the absolute density required to force a collapse is naturally much higher in the very early universe than it is towards the end of the \ac{RD} era. 

Therefore, the absolute density is not the correct metric for assessing PBH formation. Instead, the relevant physical quantity is the \emph{density contrast}, defined as the local fractional overdensity relative to the background:
\begin{equation}
    \delta \equiv \frac{\delta \rho}{\rho} = \frac{\rho_{\text{local}} - \rho}{\rho} \, .
\end{equation}
The condition for gravitational collapse to win against radiation pressure and expansion is dictated by the equation of state of the dominant fluid ($w = P/\rho$). In the \ac{RD} era ($w = 1/3$), local collapse only occurs if the density contrast exceeds a critical threshold of order unity (typically $\delta_c \sim w \approx 1/3$), regardless of how high the absolute background density $\rho$ might be. The background universe itself does not collapse because it is highly homogeneous ($\delta \sim 10^{-5}$) and expanding, perfectly balancing the average energy density~\cite{Carr:1975qj}.

\subsection{Primordial Void Rebounce Channel}

Traditionally, the literature surrounding \ac{PBH} formation has focused exclusively on the collapse of large, positive curvature perturbations—regions characterized by an extreme initial overdensity ($\delta > 0$). However, recent advancements in \ac{NR} have demonstrated that deep negative curvature perturbations, corresponding to primordial voids (PVs) where $\delta < 0$, provide a robust and entirely distinct channel for PBH formation~\cite{Joana:2025gqf}. 

For instance, Joana and Yuwen~\cite{Joana:2025gqf} have shown that deep void-like profiles in the primordial plasma do not simply disperse upon Hubble re-entry. Instead, the highly non-linear fluid dynamics of the \ac{RD} era dictate a vastly different evolution. As the underdense region expands, it drives a severe hydrodynamic shock into the surrounding background, causing the void to undergo a violent \emph{rebounce} at its center. 

This rebounce generates an effective overdensity shell. If the initial negative amplitude of the void is sufficiently deep, the mass-energy accumulated within this rebounding shell exceeds the critical threshold required for gravitational collapse, ultimately imploding to form a PBH. This rebounce mechanism fundamentally alters our understanding of the PBH mass function, as it implies that both extreme peaks and extreme troughs in the primordial density field actively contribute to the cosmological black hole abundance. The inclusion of this dual-channel formation mechanism strongly motivates the use of \emph{Peak Theory}~\cite{Bardeen:1985tr, Green:2004wb} to evaluate the PBH abundance, rather than standard Press-Schechter excursion set methods. While Press-Schechter formalism simply integrates the volume fraction of a field above a threshold, peak theory rigorously computes the distinct comoving number density of discrete spatial extrema --- allowing us to independently track and sum the populations of collapsing local maxima (standard overdensities) and collapsing local minima (voids). By expanding the statistical formation criteria within the peak-theory framework to include these non-linear void collapses, the theoretical population of PBHs generated by a given inflationary power spectrum can be significantly enhanced, providing a much more exhaustive mapping of early-universe fluctuations.

\begin{figure*}[!htb]
\centering
\includegraphics[width=0.8\textwidth]{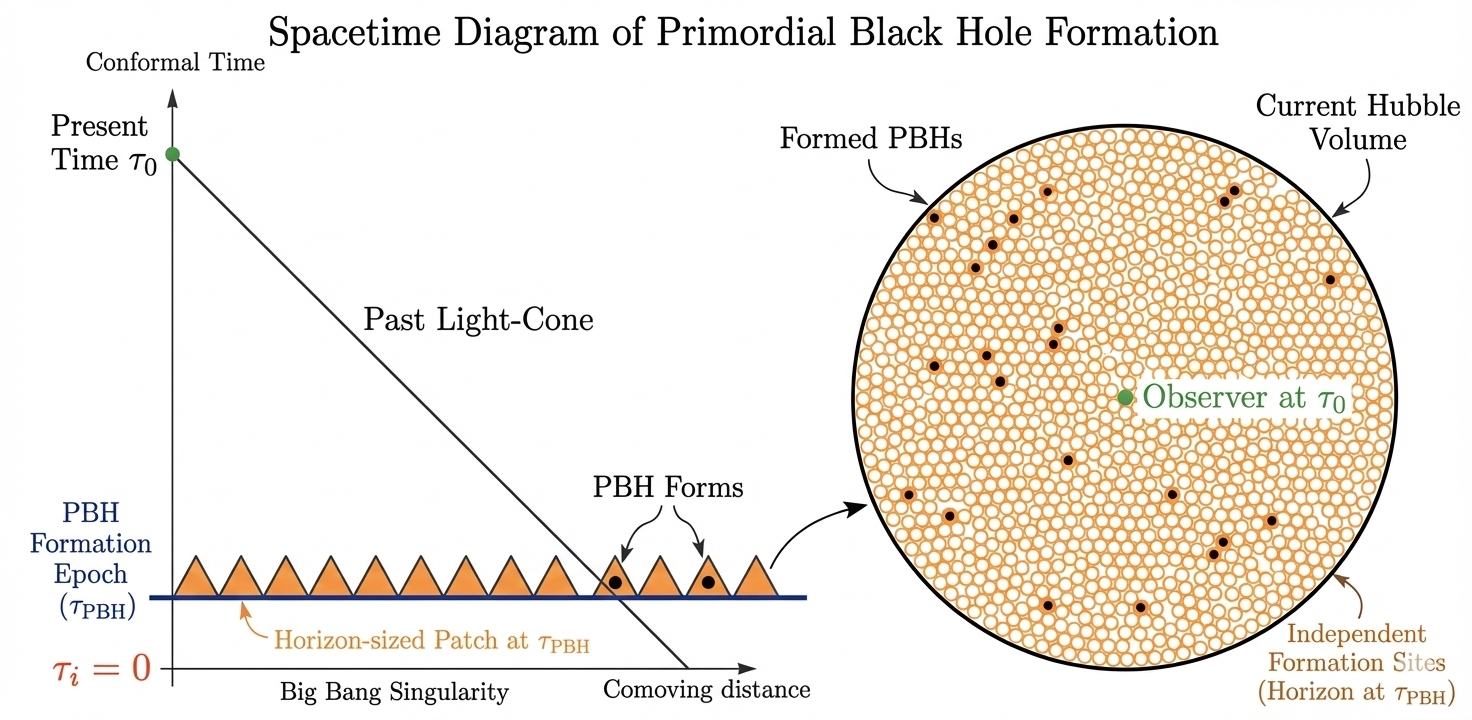} 
\caption{Space-time diagram of PBH formation during the \ac{RD} epoch within the Hubble patch. Credit: Nanobanana}
\label{fig:Space-time-PBHformation}
\end{figure*}

\section{The Number of PBH}
\label{sec:PBHNumber}

To estimate the number of \acp{PBH} existing today, we must connect the geometry of the early universe to the statistics of collapse. This requires understanding the hierarchy of causal horizons—from the macroscopic scales of the \ac{CMBR} down to the microscopic scales of PBH formation \emph{during \ac{RD} epoch}. Please note that the analysis presented below will not be applicable to the PBHs forming during (early) matter-dominated epoch.

\subsection{The Horizon Hierarchy and Causal Disconnectedness}

The homogeneity of the Universe on large scales provides the boundary condition for our counting exercise. At the last scattering surface, the particle horizon corresponds to an angular scale of $\theta \sim 1^\circ$ as seen from Earth today. No physical processes could have acted on scales larger than this at that time. 

When we observe \ac{CMBR} photons arriving from all directions, we are observing a spherical surface surrounding us at a comoving distance of $\sim 14$ Gpc ($z \approx 1100$). The remarkable isotropy of the CMBR temperature ($\Delta T/T \sim 10^{-5}$) poses the fundamental \emph{Horizon Problem} for the standard Big Bang theory: the observed sky is composed of roughly 40,000 regions that were causally disconnected at the time of recombination \cite{Liddle:2000cg}. 

This geometric picture defines our \emph{counting container}. The current Hubble patch (our observable universe) is a mixture of these smaller, ancient horizons.
\begin{description}
\item[\textbf{CMBR Scale ($t \sim 380$ kyr)}] Our patch contains $\sim 4 \times 10^4$ horizon volumes. 
 
To quantify this, we observe that a full sphere covers $4\pi$ steradians, which is equivalent to approximately $41,253$~square degrees. If we approximate each independent causal region as a 
$1^{\circ} \times 1^{\circ}$ square, thus, the total number of such regions contained within the observable sky is roughly 41,253, typically expressed as 
$\sim 4 \times 10^4$  horizon volumes.

More rigorously, this counting can be expressed as a volume ratio: 
\begin{equation} 
N_{\text{patch}}(t_{\text{rec}}) = \frac{V_0}{V_{\text{patch}}(t_{\text{rec}})} = \left( \frac{a(t_{\text{rec}})H(t_{\text{rec}})}{H_0} \right)^3 
\end{equation} 
where $V_0$ is the current comoving Hubble volume and $V_{\text{patch}}(t_{\text{rec}})$ 
is the comoving volume of a single causal horizon at recombination. Substituting standard cosmological parameters for $a(t_{\text{rec}})$  and $H(t_{\text{rec}})$ 
into this relation yields an identical order-of-magnitude result.

\item[\textbf{PBH Scale ($t \ll 1$ s)}] If we rewind further to the epoch of PBH formation, the causal horizon was microscopic. Our current universe contains a vast number of these tiny, independent \emph{formation sites.}
\end{description}
Figure~\eqref{fig:Space-time-PBHformation} provides a birds-eye view of the PBH formation within the Hubble patch.

\subsection{Counting the number of PBHs}

The total number of PBHs formed at a cosmic time $t$ during the \ac{RD} era is determined by two factors: the number of independent comoving horizon volumes contained within the currently observable Universe, and the collapse probability $\beta(t)$ for each region.

The comoving horizon scale at time $t$ is defined by the comoving Hubble radius $[a(t) H(t)]^{-1}$. Since PBH formation occurs shortly after horizon re-entry, as we saw in Sec. \eqref{sec:JeansBarrier}, there is an approximate one-to-one correspondence between the PBH mass, the comoving wavenumber $k$, and the horizon crossing time ($k=aH$).

The comoving volume of a single horizon patch at time $t$ is $V_{\rm patch}(t) \sim (a(t) H(t))^{-3}$. Given the expansion history, $V_{\rm patch} \sim t^{3/2}$ during \ac{RD} era ($a \sim t^{1/2}$) and $V_{\rm patch} \sim t$ during matter domination ($a \sim t^{2/3}$). The total comoving volume of the observable universe today is fixed by the current Hubble constant, $V_0 \sim (H_0)^{-3}$ (setting $a_0=1$).

Consequently, the number of independent horizon-sized regions (formation sites) from epoch $t$ contained within the present-day observable universe is the ratio of these volumes:
\begin{align}
\!\!\!\!\!\! 
N_{\rm patch}(t) = \frac{V_0}{V_{\rm patch}(t)} = \left[\frac{a(t) H(t)}{H_0} \right]^3  
= (a(t) E(t))^3.
\end{align}
where $E(t) = H(t)/H_0$ is the dimensionless Hubble parameter~\cite{Liu:2024-H0,Croton:2013-H0} and depends on the background cosmology. Since $N_{\rm patch} \gg 1$, this represents the total number of independent perturbations available for potential collapse, regardless of their amplitude.\footnote{The distinction between overdense and underdense regions is handled by the probability distribution; $N_{\rm patch}$ strictly counts the total number of causal domains.}

The actual number of PBHs formed is the product of these available sites and the collapse probability $\beta(t)$, which is non-zero only for regions satisfying $\delta > \delta_c$:
\begin{align}
N_{\rm PBH}(t) = N_{\rm patch}(t) \beta(t) = \left(a(t) E(t)\right)^3 \, \beta(t)~~.
\end{align}
Precise estimation requires evolving $H(t)$ (or $E(t)$) via the Friedmann equations throughout the subsequent cosmic history. By mapping the formation time $t$ to the horizon mass $M$, this relation can be expressed as a mass function:
\begin{align}
N_{\rm PBH}(M) = \left(a(t) E(t) \right)^3 \, \beta(M)~~.
\end{align}
Using observational upper limits on the abundance $\beta(M)$ (e.g., from Ref.~\cite{2021-Carr.etal-Rept.Prog.Phys}, Fig. 18), one can directly constrain the maximum number of PBHs, $N_{\rm PBH}(M)$, formed at any specific mass scale.

\subsection{Counting the Formation Sites}

We treat the early universe as a collection of independent Hubble volumes, each representing a potential site for PBH formation. Let $M_{H,0} \approx 10^{22} M_{\odot}$ be the mass of the current Hubble horizon. As discussed in Sec.~\eqref{sec:scale_invariance}, a PBH forming at time $t_{\rm form}$ has a mass roughly equal to the horizon mass at that time, $M_{\text{PBH}} \sim M_H(t_{\rm form})$ \cite{Carr:1975qj}.

The total number of independent formation sites ($N_{\text{sites}}$) available within our current observable universe is the ratio of the total mass to the formation mass:
\begin{equation}
    \label{eq:Nsites}
    N_{\text{sites}}(M) \approx \frac{M_{H,0}}{M_{\text{PBH}}} \, .
\end{equation}
For a typical PBH mass of $M_{\text{PBH}} \sim 10^{15}$ g ($\sim 10^{-18} M_{\odot}$), the number of sites is astronomical:
\begin{equation}
    N_{\text{sites}} \approx \frac{10^{22} M_{\odot}}{10^{-18} M_{\odot}} = 10^{40} \, .
\end{equation}
This indicates that the early universe comprised approximately $10^{40}$ independent causal regions, each representing a potential site for gravitational collapse. 

\subsection{The Collapse Probability (\texorpdfstring{$\beta$}{beta})} 

While the number of sites is huge, the probability of any single site collapsing is governed by the amplitude of density fluctuations. The abundance of PBHs is defined by the mass fraction $\beta(M)$:
\begin{equation}
    \beta(M) \equiv \frac{\rho_{\text{PBH}}(t_{\rm form})}{\rho_{\text{tot}}(t_{\rm form})} = \int_{\delta_c}^1 P(\delta) \, d\delta \, .
\end{equation}
{Often, a multiplicative factor `2' is included in the second equality on the right-hand side. This originates from the naive application of the original Press-Schechter formalism~\cite{Press:1973iz} proposed for the halo formation to the case of PBH. However, recent studies based on the excursion set formalism for PBH formation during the \ac{RD} epoch show that this multiplicative factor is not required~\cite{Kushwaha:2025zpz,Saito:2025sny}.
Assuming the density contrast $\delta$ follows a Gaussian probability distribution $P(\delta)$, and using the high-threshold approximation ($\delta_c \gg \sigma$), we obtain:
\begin{equation}
\label{eq:Threshold-Beta}
    \beta(M) \approx \frac{1}{2}\text{erfc}\left( \frac{\delta_c}{\sqrt{2}\sigma(M)} \right) \approx \frac{\sigma(M)}{\sqrt{2\pi}\delta_c} \exp\left( -\frac{\delta_c^2}{2\sigma^2(M)} \right) \, .
\end{equation}
This resolution highlights that while $\sigma \sim 10^{-5}$ on CMBR scales (leading to $\beta \to 0$), the power spectrum on small scales must be significantly enhanced for $\sigma$ to approach the threshold $\delta_c \sim 1/3$ and produce a non-negligible abundance.\\
}

\subsection{Total Number of PBHs: Formation vs. Survival}

To estimate the total population of PBHs, one must distinguish between the number of objects \emph{formed} in the early universe and the number surviving \emph{today}.

The standard cosmological parameter $\Omega_{\text{PBH}}$ describes the current density of PBHs relative to the critical density. However, this parameter is only valid for PBHs that have survived to the present day (i.e., those with $M > M_* \approx 5 \times 10^{14}$ g). (Detailed discussion in Sec. \eqref{sec:HawkingRad}.) 
For lighter BHs that have already evaporated, $\Omega_{\text{PBH}}$ is effectively zero, yet their formation number was non-zero and physically significant.
Therefore, the fundamental parameter for counting is the initial mass fraction $\beta(M)$, which quantifies the production rate at the epoch of formation $t_{\rm form}$, regardless of future evaporation. 

The total number of PBHs formed within our comoving volume is the product of the available formation sites and the collapse probability, enhanced by the differential expansion of matter vs. radiation. Since PBHs behave as dust ($\rho \propto a^{-3}$) while the background radiation scales as $\rho \propto a^{-4}$, the PBH number density relative to photons is fixed at formation but the energy density ratio grows linearly with the scale factor.

The total number of PBHs formed at a specific mass scale $M$ is given by \cite{Sasaki:2018dmp}:
\begin{equation}
    N_{\text{formed}}^{\text{(PBH)}}(M) \approx \beta(M) \times N_{\text{sites}}(M) \times \left( \frac{a(t_{eq})}{a(t_f)} \right) \, .
\end{equation}
This relation holds for all mass scales. For massive PBHs ($M > 10^{15}$ g), this number corresponds to the \ac{DM} population observable today. For microscopic PBHs ($M < 10^{15}$ g), this number represents the total population that existed prior to evaporation, whose integrated Hawking radiation may still leave detectable imprints on Big Bang Nucleosynthesis (BBN) or the \ac{CMBR}.

\subsection{From $N_{\text{formed}}^{\text{(PBH)}}$ to $\Omega_{\text{PBH}}$}
\label{sec:AbundanceDerivation}

We can now derive the PBH abundance constraint ($\Omega_{\text{PBH}}$) directly from the counting argument established above. This step bridges the gap between the discrete number of formation events, $N_{\text{formed}}$, and the continuous density parameter used in cosmological constraints.

For PBHs massive enough to survive Hawking evaporation ($M \gtrsim 5 \times 10^{14}$ g), the initial mass fraction $\beta$ translates directly into a contribution to the current \ac{DM} density. By definition, $\beta$ is the ratio of PBH density to the total (radiation) density at the precise moment of formation $t_{\rm form}$:
\begin{equation}
\label{def:beta-form}
    \beta(M) = \frac{\rho_{\text{PBH}}(t_{\rm form})}{\rho_{\text{tot}}(t_{\rm form})} \, .
\end{equation}
As the universe expands, the radiation density dilutes as $\rho_{\text{rad}} \propto a^{-4}$, whereas PBHs (behaving as non-relativistic matter) dilute as $\rho_{\text{PBH}} \propto a^{-3}$. Consequently, the density contrast of the black hole population grows linearly with the scale factor $a$:
\begin{equation}
\label{eq:rhoPBH-rad}
    \frac{\rho_{\text{PBH}}(t)}{\rho_{\text{rad}}(t)} = \beta \frac{a(t)}{a(t_{\rm form})} \, .
\end{equation}
This linear growth continues until the epoch of matter-radiation equality ($t_{\text{eq}}$). At matter-radiation equality $t_{\text{eq}}$, the above ratio becomes unity (by definition of equality, assuming PBHs make up all DM for the moment, or scaling with $\Omega_{\text{DM}}$):
\begin{equation}
\label{eq:rhoPBH-rad-equality}
    \frac{\rho_{\text{PBH}}(t_{\text{eq}})}{\rho_{\text{rad}}(t_{\text{eq}})} \approx \frac{\Omega_{\text{PBH}}}{\Omega_{\text{rad}}} \frac{a(t_{\text{eq}})}{a(t_0)} = \frac{\Omega_{\text{PBH}}}{\Omega_{\text{rad}}} \frac{1}{(1+z_{\text{eq}})} \, .
\end{equation}
where $z$ represents cosmological redshift. The current density parameter $\Omega_{\text{PBH}}$ is determined by projecting this ratio to today. Assuming standard adiabatic expansion ($g_{*s} T^3 a^3 = \text{const}$), the scale factor ratio relates to the temperature ratio:
\begin{equation}
 \frac{a(t_{\text{eq}})}{a(t_{\rm form})} \approx \frac{T_{\rm form}}{T_{\text{eq}}} \left( \frac{g_{*s}(t_{\rm form})}{g_{*s}(t_{\text{eq}})} \right)^{1/3} \, .
\end{equation}
The ratio of the scale factor today ($a_0$) to the scale factor at formation ($a_{\rm form}$) is:
\begin{equation}
\frac{a_0}{a_{\rm form}} = \frac{T_{\rm form}}{T_0} \left( \frac{g_{*s}(t_{\rm form})}{g_{*s}(t_0)} \right)^{1/3} \, .
\end{equation}
To recover the dependence on mass $M$, we recall that the formation mass corresponds to the horizon mass: 
\begin{equation}
    M = \gamma M_H = \gamma \frac{4\pi}{3} \rho_{\text{rad}} H_f^{-3} \approx \gamma \frac{c^3}{2 G H_f} \, .
\end{equation}
Using the Friedmann equation $H^2 = \frac{8\pi G}{3} \rho_{\text{rad}}$ and the radiation density $\rho_{\text{rad}} = \frac{\pi^2}{30} g_* T^4$, we can solve for $T_{\rm form}$:
\begin{equation}
    k_B T_{\rm form} \approx \left( \frac{45 \hbar^3 c^5}{16 \pi^3 G \gamma^2 g_*(T_{\rm form})} \right)^{1/4} M^{-1/2} \, .
\end{equation}
Substituting standard numerical values ($g_{*s}(t_0) = 3.91$, $T_0 = 2.725$ K, $M_{\odot} \approx 1.99 \times 10^{33}$ g), we can express the scale factor ratio in terms of the solar mass. The current density is then:
\begin{eqnarray}
\Omega_{\text{PBH}} &=& \frac{\rho_{\text{PBH}}(t_0)}{\rho_c} = \Omega_{\text{rad}} \frac{\rho_{\text{PBH}}(t_0)}{\rho_{\text{rad}}(t_0)} \nonumber \\
&=&  \beta \, \Omega_{\text{rad}} \frac{a_0}{a_{\rm form}} 
\left( \frac{g_{*s}(t_{\rm form})}{g_{*s}(t_0)} \right) \, .
\end{eqnarray}
Combining the temperature-mass relation with the entropy conservation factor yields the final expression~\cite{Carr:2009jm}:
\begin{eqnarray}
    \Omega_{\text{PBH}} h^2 &\approx& 2.7 \times 10^8 \left( \frac{\gamma}{0.2} \right)^{1/2} \left( \frac{g_*(t_{\rm form})}{106.75} \right)^{-1/3} \\ \nonumber
  & \times &  \left( \frac{M}{M_{\odot}} \right)^{-1/2} \beta(M) \, .
\end{eqnarray}

The relation derived above reveals fine-tuning. The pre-factor of $\sim 10^8$ arises from the immense expansion between formation ($t_{\rm form}$) and equality ($t_{\text{eq}}$). Because PBHs behave as dust ($\rho \propto a^{-3}$) in a \ac{RD} universe ($\rho \propto a^{-4}$), their relative density is amplified by eight orders of magnitude during this epoch. To end up with the correct amount of \ac{DM} today ($\Omega_{\text{PBH}} \approx 0.26$), the initial abundance must have been vanishingly small, $\beta(M_{\odot}) \sim 10^{-9}$. This result underscores that PBH formation is not a generic feature of the early universe, but a rare phenomenon restricted to roughly one part per billion of the total cosmic volume at that epoch.

\subsection{Thermodynamic Consistency and Number Density}
\label{sec:Thermo}

To validate the geometric counting of formation sites, we must ensure it is consistent with the thermodynamic evolution of the universe. Our site-counting estimation yields a proportionality of $N_{\text{formed}} \propto \beta M^{-3/2}$ [cf. Eq.~\eqref{eq:Threshold-Beta}]. This $M^{-3/2}$ scaling physically signifies that while fewer independent horizon-sized sites exist for larger masses ($\sim M^{-1}$), their relative density contrast grows for longer periods during the \ac{RD} era ($\sim M^{1/2}$), resulting in a characteristic mass spectrum. 

Evaluation at the asteroid-mass scale ($M \sim 10^{20}$ g) confirms that $N_{\text{sites}} \sim 10^{37}$ [cf. Eq.~\eqref{eq:Nsites}], reinforcing the conclusion that PBH formation is an exceptionally rare fluctuation in the high-energy tail of the primordial density field~\cite{Carr:1975qj}. As we will rigorously derive in Sec.~(\ref{sec:Abundance}), this geometric scaling perfectly matches the exact thermodynamic calculations for the present-day comoving number density $n_{\text{PBH}}(t_0)$ assuming adiabatic expansion and entropy conservation~\cite{Carr:2021bzv}.

\section{The Rigorous Recipe: Compaction and Criticality}
\label{sec:Compaction}

Until this point, our analysis has relied primarily on the kinematics of the background FLRW geometry, without requiring the explicit solution of the non-linear dynamical equations. We now transition to the exact relativistic calculations of gravitational collapse.

Independent of the specific primordial mechanism that generates density fluctuations --- whether from standard inflation, multi-field dynamics, or phase transitions --- the fundamental relativistic question remains: \textit{What exact geometric condition dictates the gravitational collapse of a local overdensity into a \ac{BH}?}

Historically, the criterion for collapse was approximated using the density contrast $\delta(t, \textbf{x})$~\cite{Carr:1974nx,Carr:1975qj,Carr:1976zz}. However, $\delta(t, \textbf{x})$ is an inherently gauge-dependent quantity (see discussion in Appendix \eqref{app:JeansDetail}), leading to physical ambiguities when evaluated on super-horizon scales. To resolve this, the modern standard, developed extensively through \ac{NR}, abandons the density contrast $\delta(t, \textbf{x})$ in favor of a robust, gauge-invariant geometric quantity \emph{Compaction Function}, $\mathcal{C}$ defined in Sec. \eqref{sec:Hoop}. 

The transition from a primordial density perturbation to a BH is best understood through the rigorous geometric framework outlined below.

\subsection{Super-horizon geometry and Misner-Sharp mass}
\label{sec:SHorizon-MSharpMass}

We consider the formation of a PBH originating from a large, isolated curvature fluctuation. Since the characteristic scale of the perturbation is initially much larger than the cosmological horizon, the initial conditions can be robustly established using the \emph{gradient expansion} (or long-wavelength) approximation~\cite{Wands:2000dp,Lyth:2004gb,Tanaka:2006zp}. 

On super-horizon scales, the physical size of the perturbation is much larger than the Hubble radius, meaning the characteristic comoving wavenumber $k$ satisfies $k \ll aH$. Consequently, we can define a small perturbative expansion parameter $\epsilon \equiv k/aH\ll 1$. In this regime, spatial gradients are heavily suppressed relative to time derivatives, making the gradient expansion an exceptionally accurate mathematical tool to track the non-linear metric. \footnote{For completeness we have provided a brief discussion in Appendix~\eqref{sec:GradientExpansion}. Appendix \eqref{sec:NewtonianRelativisticLink} contains the connection between the Newtonian derivation and relativistic gradient expansion derivation.}  

Within this super-horizon regime, the spherically symmetric spatial metric is parameterized directly by the non-linear curvature perturbation $\zeta(r)$~\cite{Shibata:1999zs,Polnarev:2006aa}:
\begin{equation}
\label{eq:NLCurv-FLRW}
ds^2 = -dt^2 + a^2(t) e^{2\zeta(r)} \left( dr^2 + r^2 d\Omega^2 \right) \, ,
\end{equation}
where $a(t)$ is the background cosmological scale factor, $r$ is the comoving radial coordinate, and $\zeta(r)$ acts as a gauge-invariant measure of the spatial curvature~\cite{Wands:2000dp,Lyth:2004gb}. 

Before proceeding further, it is crucial to emphasize the strict domain of validity for this approximation: it must only be applied prior to horizon crossing. As the perturbation approaches and re-enters the cosmological horizon ($k \gtrsim aH$), the expansion parameter grows ($\epsilon \gtrsim 1$), causing the spatial derivative series to formally diverge. Therefore, the gradient expansion is used exclusively to establish the exact, time-independent initial conditions of the perturbation (such as the compaction function), whereas the subsequent dynamical collapse upon re-entry must be modeled using full \ac{NR} or realistic fluid solutions \cite{Polnarev:2006aa, Harada:2013epa}. 

To assess whether this geometry will ultimately collapse, one must determine the mass enclosed within a given radius. As discussed in Sec. \eqref{sec:MSDef-Kodama}, in spherical symmetry, the unique and rigorously defined measure of quasi-local energy is the \emph{Misner-Sharp mass}, $M_{MS}$. To compute it, we first define the physical, or \emph{Areal Radius}, $R(r,t)$~\cite{Shibata:1999zs}:
\begin{equation}
R(r,t) = a(t) r e^{\zeta(r)} \, .
\end{equation}
The Misner-Sharp mass is then defined geometrically by the gradient of this areal radius~\cite{1964-Misner_Sharp-PRD,Abreu:2010ru}:
\begin{equation}
M_{MS}(r,t) = \frac{R}{2G} \left( 1 - g^{\mu\nu} \partial_\mu R \partial_\nu R \right) \, .
\label{def:MSharp}
\end{equation}
Because we are working in the super-horizon limit, the adiabatic condition dictates that the curvature perturbation is frozen, meaning $\zeta$ depends exclusively on the radial coordinate $r$. Therefore, the temporal derivative is simply driven by the background expansion, $\dot{R} = \dot{a}re^{\zeta(r)} = HR$, while the spatial derivative is $R' = \frac{R}{r}(1 + r\zeta'(r))$. Substituting these into the inverse metric ($g^{tt} = -1$, $g^{rr} = a^{-2}e^{-2\zeta} = r^2/R^2$), Eq.~\eqref{def:MSharp} becomes exactly:
\begin{equation}
    M_{MS}(r,t) = \frac{R}{2G} \left[ 1 + \dot{R}^2 - (1 + r\zeta'(r))^2 \right] \, .
\end{equation}
The total enclosed mass is heavily dominated by the background FLRW density, which is characterized by the expansion term $M_{bg} = \frac{R}{2G}\dot{R}^2$. To isolate the physics of the collapse, we must define the \emph{mass excess}, $\delta M$, which removes this dominating background expansion:
\begin{eqnarray}
\delta M(r,t) &\equiv& 
M_{MS}(r,t) - M_{bg}(r,t) \nonumber \\
&=& \frac{R}{2G} \left[ 1 - (1 + r\zeta'(r))^2 \right] \simeq - \frac{r R}{G} \zeta'\, .
\end{eqnarray}
Thus, we see that the enclosed mass excess driving PBH formation manifests entirely as a consequence of the local spatial curvature gradient, $\zeta'(r)$.

\subsection{The Compaction Function}

While the Misner-Sharp mass provides the total energy within a volume, collapse is not driven by the absolute mass, which includes the background cosmological density. Rather, it is driven by the \emph{excess mass}. To isolate this gravitational pull, we define the {Compaction Function} ($\mathcal{C}(r)$), elaborated in Sec.~\eqref{sec:Hoop}, as twice the ratio of the excess mass to the areal radius~\cite{Shibata:1999zs,Harada:2013epa}:
\begin{equation}
\mathcal{C}(r) \equiv \frac{2 G \, \delta M(r)}{R(r)} \, .
\end{equation}
As shown in Appendix \eqref{sec:DerivationDelta}, by substituting the metric components into the Misner-Sharp Mass definition \eqref{def:MSharp}, and subtracting the background FLRW energy density, the compaction function can be expressed purely in terms of the primordial curvature profile $\zeta(r)$. Crucially, to evaluate the compaction function accurately at the epoch of horizon crossing, one must incorporate the linear transfer function that relates the super-horizon curvature to the density field. As rigorously shown by Musco~\cite{Musco:2018rwt}, this introduces an equation-of-state dependent factor, yielding:
\begin{equation}
\label{eq:CompactionZeta}
\mathcal{C}(r) = -\frac{6(1+w)}{5+3w} r \frac{d \zeta(r)}{dr} 
\left[ 1 + \frac{r}{2} \frac{d \zeta(r)}{dr} \right] \, ,
\end{equation}
where $w$ is the equation of state parameter of the cosmological fluid. (For a standard \ac{RD} epoch, $w=1/3$, the pre-factor simplifies exactly to $-4/3$).

As detailed in Sec.~\eqref{sec:Hoop}, the compaction function $\mathcal{C}(r)$ acts as a relativistic generalization of the Hoop Conjecture. It measures the \emph{absolute depth} of the local gravitational potential well. Crucially, on super-horizon scales, $\mathcal{C}(r)$ is time-independent; it is an initial condition locked into the shape of the profile $\zeta(r)$ long before the perturbation re-enters the horizon. Details are provided in Appendix \eqref{sec:GradientExpansion}.

\subsection{Universal Thresholds and Critical Scaling}

Thus, the criterion for \ac{BH} formation translates smoothly into a question of whether the potential well is sufficiently deep to trap light. We characterize the entire perturbation by the maximum amplitude of its compaction function, $\mathcal{C}_{\max}$, located at some characteristic comoving radius $r_m$. A PBH will form if and only if this peak amplitude exceeds a critical threshold~\cite{Musco:2018rwt,Escriva:2019phb}:
\begin{equation}
\mathcal{C}_{\max} > \mathcal{C}_c \, .
\end{equation}
Extensive \ac{NR} simulations have precisely mapped this critical threshold $\mathcal{C}_c$. For \ac{RD} equation of state ($w=1/3$), the threshold is found to be~\cite{Musco:2018rwt,Escriva:2019phb}:
\begin{equation}
\mathcal{C}_c \approx 0.4 - 0.5 \, .
\end{equation}
While the exact value carries a mild dependence on the specific geometric shape of the profile $\zeta(r)$, a value of $\sim 0.45$ serves as the canonical standard. It is important to emphasize that this critical threshold is highly sensitive to the background equation of state parameter, $w$, at the exact epoch of horizon reentry. A comprehensive analysis of how $\mathcal{C}_c$ scales with softer or stiffer equations of state --- such as during an early matter-dominated phase or a QCD phase transition --- is detailed in Appendix~(\ref{sec:TopHatDerivation}).

If this threshold is breached, the gravitational collapse proceeds. However, the final state is not a single, universally sized black hole. Instead, the \emph{ideal formation} channel obeys the laws of \emph{Critical Collapse}, originally discovered by Choptuik~\cite{Choptuik:1992jv,Gundlach:2007gc}. The mass of the resulting PBH is deterministically linked to how far $\mathcal{C}_{\max}$ exceeds the threshold, following a universal scaling law~\cite{Niemeyer:1997mt,Gow:2020bzo}:
\begin{equation}
M_{PBH} = K M_H(t_H) \left( \mathcal{C}_{\max} - \mathcal{C}_c \right)^{\upsilon_{\rm c}} \, .
\end{equation}
Here, $M_H(t_H)$ represents the mass of the cosmological horizon at the moment the perturbation re-enters; $\upsilon$ is a universal critical exponent determined entirely by the background fluid (with $\upsilon_{\rm c} \approx 0.36$ for radiation); and $K$ is an order-unity constant dependent on the initial profile shape. 

Ultimately, the modern relativistic recipe for PBH formation is remarkably elegant: it maps a primordial curvature profile $\zeta(r)$ to its compaction function $\mathcal{C}(r)$ and identifying the peak $\mathcal{C}_{\max}$. If this peak exceeds the critical threshold $\mathcal{C}_c$, a black hole is born, with a final mass dictated completely by the critical scaling law. This formulation strips away the historical ambiguities of the density contrast, providing a rigorous, gauge-invariant approach from quantum fluctuations to macroscopic BHs.

\subsection{Detailed derivation from super-horizon metric}
\label{sec:DetailedCompaction}

To understand precisely how the compaction function is constructed, we evaluate the Misner-Sharp mass directly using the gradient expansion metric derived above. 

In the typical regime where the curvature perturbation is small ($\zeta \ll 1$), 
we can linearize \eqref{eq:CompactionZeta} leading to:
\begin{equation}
\label{eq:CompactionLinear}
\mathcal{C}(r) \approx - (4/3) \, r \zeta'(r) \, .
\end{equation}

From the above expression we can understand the mathematical superiority of $\mathcal{C}(r)$ over the density contrast $\delta(r)$ is two different ways: 
\begin{enumerate}
    \item \textbf{Insensitivity to Constants:} 
    Since it depends exclusively on the spatial gradient $\zeta'$, adding a constant shift $\zeta \to \zeta + C$ (which corresponds to a global change in coordinates or a \emph{separate universe} scaling) leaves $\mathcal{C}$ entirely unchanged.\footnote{Regarding other background Killing symmetries of the metric in Eq.~\eqref{eq:NLCurv-FLRW}: spatial rotations trivially annihilate the spherically symmetric profile ($\partial_\theta \zeta = \partial_\phi \zeta = 0$), while spatial translations merely shift the coordinate center of the collapse. The shift $\zeta \to \zeta + C$ is therefore unique, as it is tied directly to the dilatation symmetry of the spatial slice, representing a residual large gauge degree of freedom.} Details in Appendix \eqref{sec:DerivationDelta}.
    \item \textbf{Integrated Smoothing:} The density contrast, besides being a gauge-dependent quantity, is also proportional to the Laplacian of the curvature ($\delta \propto \nabla^2 \zeta$), making it a local point function highly sensitive to small-scale noise. Conversely, $\mathcal{C} \propto \zeta'$ is proportional to the volume integral of $\delta$. Specifically, $\mathcal{C}$ evaluated at a radius $R$ reflects the \emph{mean density} enclosed within that volume:
    \begin{equation}
    \mathcal{C}(R) \sim \frac{1}{R} \int_0^R \delta(r) \, r^2 \, dr \, .
    \end{equation}
    This integration acts as a natural physical filter, making $\mathcal{C}(r)$ a far more reliable predictor of bulk gravitational collapse than the raw, unintegrated density profile.
\end{enumerate}

\subsection{Physical intuition: The pressure-gradient penalty}
\label{sec:TopHatVsRealistic}

To understand why the numerically determined threshold for $\mathcal{C}_c$ is higher than simple analytic estimates, it is instructive to revisit the idealized \emph{Top-Hat density profile}. 

For a uniform sphere of radius $R$ and excess density $\delta \rho = \bar{\rho} \delta$, the mass excess is $\delta M = (4\pi/3) R^3 \bar{\rho} \delta$. Substituting this into the compaction function definition (cf. Eq.~\eqref{eq:CompactionZeta}) yields:
\begin{equation}
\mathcal{C} = \frac{2G}{R} \left( \frac{4\pi}{3} R^3 \bar{\rho} \delta \right) = \left( \frac{8\pi G \bar{\rho}}{3} \right) R^2 \delta = (HR)^2 \delta \, .
\end{equation}
At the precise moment of \emph{horizon crossing}, the physical size of the perturbation matches the Hubble radius ($HR = 1$). Consequently, for a perfectly uniform sphere, the compaction function perfectly tracks the density contrast:
\begin{equation}
\mathcal{C}_{\text{Top-Hat}} = \delta_{\text{Top-Hat}} \, .
\end{equation}
Using the analytic Jeans criterion threshold derived earlier ($\delta_{\text{th}} \approx c_s^2 = 1/3$), the corresponding compaction threshold is naturally $\mathcal{C}_c^{\text{Top-Hat}} \approx 0.33$.

However, this Top-Hat model contains a drastic, unphysical simplification: the density is uniform everywhere inside the sphere. As a result, the pressure ($P = w\rho$) is also uniform, meaning there are no internal pressure gradients ($\nabla P = 0$) pushing the fluid apart prior to collapse. 

In a realistic cosmological scenario, the density profile must be smooth, typically peaking at the center and decaying radially (e.g., a Gaussian profile). Since the density is highest at the center, the pressure is also highest at the center. This establishes a powerful, outward-directed \emph{Pressure Gradient Force} ($-\nabla P$) that actively resists compression. 

To form a PBH from a realistic profile, gravity must not only halt the background Hubble expansion (which requires $\mathcal{C} \approx 0.33$), but it must \emph{additionally} overpower this internal pressure gradient. Therefore, the true threshold is inevitably pushed higher:
\begin{equation}
\mathcal{C}_c^{\text{Realistic}} = \mathcal{C}_c^{\text{Top-Hat}} + \text{Gradient Penalty} \, .
\end{equation}
For a \ac{RD} universe ($w=1/3$), full \ac{NR} simulations quantify this penalty. It is important to note that arriving at this quantitative limit assumes a sufficiently large peak density contrast ($\delta_{\text{max}} \gtrsim 1$) at the center of the perturbation, a standard prerequisite for the rare, extreme fluctuations capable of undergoing collapse. Under this condition, the simulations establish a robust, universal threshold range of $\mathcal{C}_c \approx 0.4 - 0.5$. Physically, this dictates that if the excess mass inside the horizon exceeds roughly $40\%$ to $50\%$ of the total horizon mass, gravity definitively wins the \emph{race against sound}, and a \ac{PBH} is born.

\begin{figure}[!htb]
\centering
\includegraphics[width=0.5\textwidth]{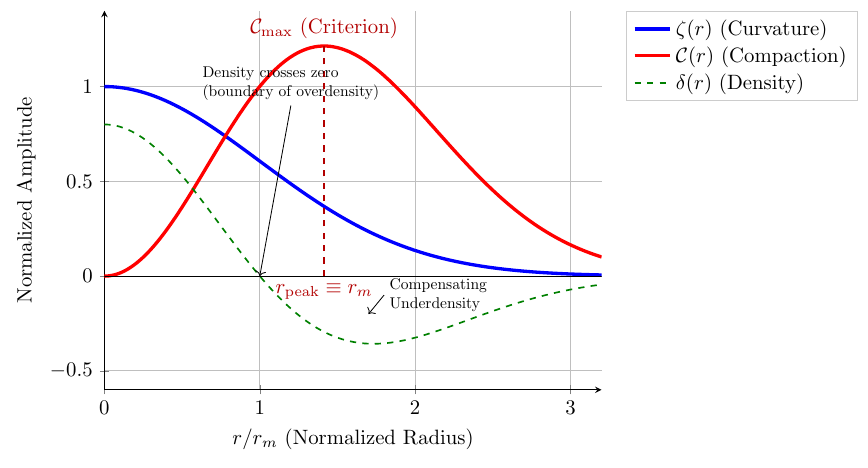} 
\caption{A schematic comparison of overdensity $\delta(r)$, compaction function $\mathcal{C}(r)$, and curvature perturbation $\zeta(r)$, illustrating how $\mathcal{C}(r)$ acts as a smoothed, integrated measure of the perturbation's gravitational depth.}
\label{fig:comparisonplot}
\end{figure}

The physical significance of this relationship is  illustrated in Fig.~(\ref{fig:comparisonplot}). The plot compares the normalized spatial profiles of the underlying curvature $\zeta(r)$, the local density contrast $\delta(r)$, and the integrated compaction function $\mathcal{C}(r)$. 
As shown, the density contrast $\delta(r)$ is sharply peaked at the origin and rapidly drops, inevitably crossing zero to form a compensating underdensity. This zero-crossing is a necessary boundary condition for isolated perturbations, ensuring they do not overclose the background universe. However, because of this rapid spatial variation, the raw local value of $\delta(r)$ does not cleanly communicate the total gravitational pull of the region.

Conversely, the compaction function $\mathcal{C}(r)$ acts as a cumulative measure. It rises smoothly from the origin, aggregating the local density variations, until it reaches a well-defined global maximum, $\mathcal{C}_{\max}$, at a characteristic comoving radius $r_m$ (defined mathematically by $\left. \mathcal{C}'(r)\right|_{r=r_m}=0$). 

This peak establishes an important mathematical equivalence between the two formalisms: if we decompose the density contrast on super-horizon scales into its time-dependent and time-independent spatial components, $\delta(t,r) = \Xi(t) \delta(r)$, it can be shown that the compaction function is precisely equal to the time-independent component of the volume-averaged density contrast, smoothed exactly up to the peak radius $r_m$. Thus, $\mathcal{C}_{\max}$ captures the true, integrated average overdensity of the gravitationally bound region, filtering out arbitrary local spikes while identifying the precise mass enclosed within $r_m$ that will ultimately decouple from the Hubble flow and collapse into a black hole.

\section{PBH Abundance and the Mass Fraction}
\label{sec:Abundance}

Once the formation criterion is established, the next task is to quantify the number density of PBHs formed. The fundamental quantity of interest is the \emph{mass fraction} ($\beta(M)$) at formation defined in Eq.~\eqref{def:beta-form}
This represents the fraction of the total energy density of the universe that collapses into \acp{BH} at the formation time $t_f$.

\subsection{The Press-Schechter Formalism}
Since the initial curvature perturbations $\zeta$ are stochastic quantum fluctuations generated during inflation, we must treat the density contrast $\delta$ as a random variable. Assuming the perturbations follow a Gaussian distribution $P(\delta)$ (as predicted by standard slow-roll inflation), the probability density function is:
\begin{equation}
P(\delta) = \frac{1}{\sqrt{2\pi}\sigma(M)} \exp\left( - \frac{\delta^2}{2\sigma^2(M)} \right) \, ,
\end{equation}
where $\sigma^2(M)$ is the variance of the density contrast smoothed on the horizon scale $R_H(M)$. This variance is directly related to the Primordial Power Spectrum $\mathcal{P}_\zeta(k)$.

A PBH forms in any region where the coarse-grained density contrast exceeds the critical threshold $\delta_c$. Strictly speaking, the upper limit of this integration is not infinity, but a maximum physical overdensity $\delta_{\text{max}}$ (corresponding to a compaction function $\mathcal{C}_{\max} \approx 2/3$)~\cite{Harada:2023ffo,Harada:2024trx}. Perturbations with $\delta > \delta_{\text{max}}$ correspond to extreme spatial curvatures, as discussed in Sec.~\eqref{sec:Pitfalls}, classified as Type II fluctuations. For Gaussian primordial fluctuations, the relative contribution from Type II fluctuations is statistically suppressed. Furthermore, recent numerical simulations demonstrate that Type II fluctuations do not necessarily result in Type II PBHs during the \ac{RD} epoch~\cite{Uehara:2024yyp}. Thus, the mass fraction is the integral of the probability tail up to this maximum cut-off:
\begin{equation}\label{beta-ps}
\beta(M) = \int_{\delta_c}^{\delta_{\text{max}}} P(\delta) \, d\delta \approx \frac{1}{2} \text{erfc}\left( \frac{\delta_c}{\sqrt{2}\sigma(M)} \right) \, .
\end{equation}
Because the Gaussian tail drops sharply, the error function remains an excellent approximation. Using the asymptotic expansion of the complementary error function ($\text{erfc}(x) \approx \frac{e^{-x^2}}{x\sqrt{\pi}}$ for large $x$), we obtain the standard analytic form:
\begin{equation}
\label{eq:beta_approx}
\beta(M) \approx \frac{\sigma(M)}{\sqrt{2\pi}\delta_c} \exp\left( - \frac{\delta_c^2}{2\sigma^2(M)} \right) \, .
\end{equation}

This exponential dependence highlights a fundamental vulnerability within the Press-Schechter approach to PBH formation: the initial mass fraction $\beta(M)$ is acutely sensitive to minor variations in the critical density threshold $\delta_c$. Because PBH formation occurs far down the tail of a Gaussian distribution, small theoretical or systematic ambiguities in pinning down the precise value of $\delta_c$ generate massive, non-linear uncertainties in the calculated abundance. 

\begin{figure}[!htb]
\centering
\includegraphics[width=0.5\textwidth]{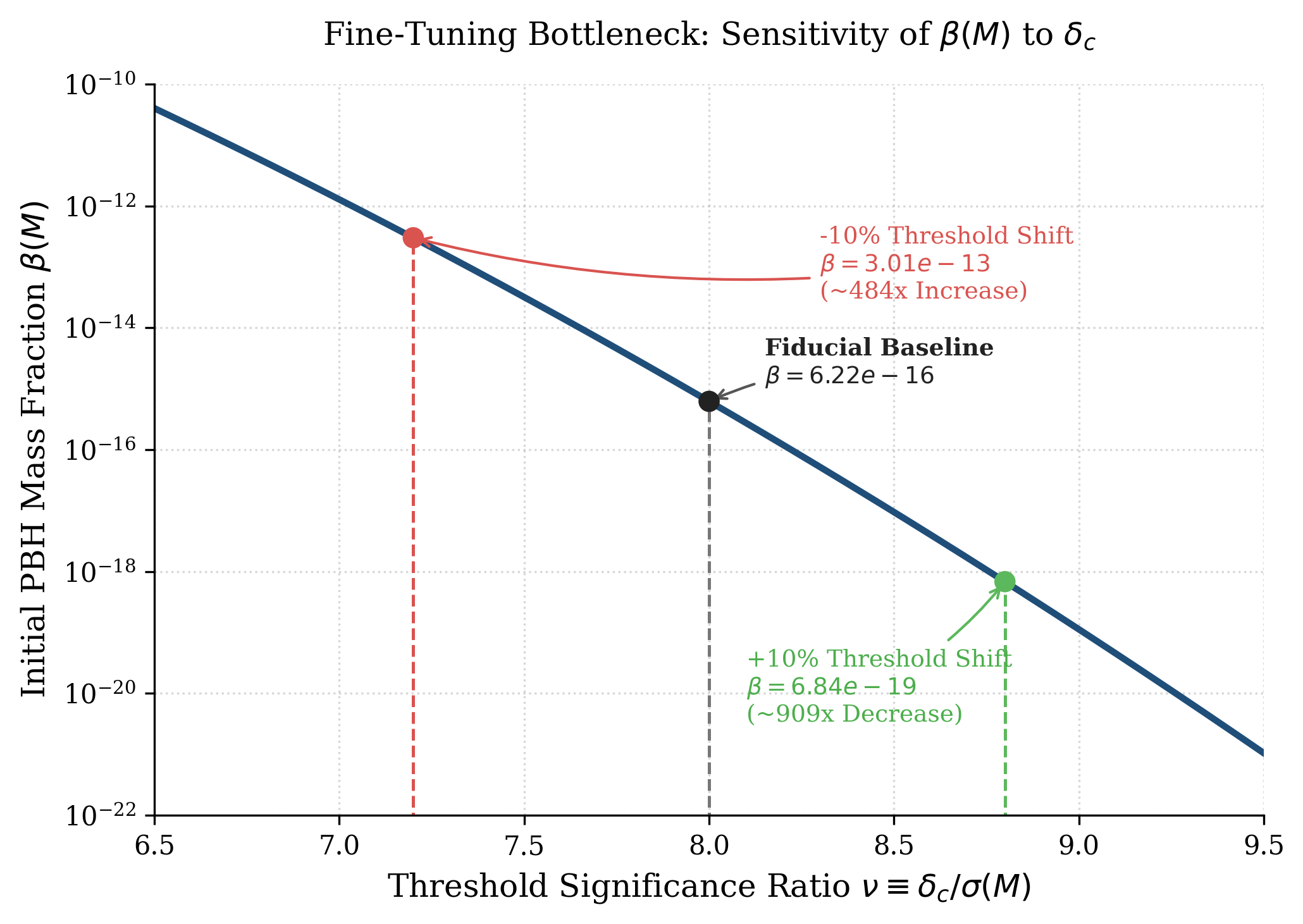} 
\caption{The extreme exponential sensitivity of the primordial black hole collapse fraction $\beta(M)$ relative to minor shifts in the critical threshold ratio $\nu \equiv \delta_c / \sigma(M)$. Taking a standard dark matter baseline at $\nu = 8.0$ ($\beta \approx 6.22 \times 10^{-16}$), a minute $-10\%$ drop in the evaluation of $\delta_c$ inflates the black hole abundance by a factor of $\sim 484$, whereas a $+10\%$ structural increase drops the resulting abundance by nearly three orders of magnitude ($\sim 910$ times lower).}
    \label{fig:pbh_fine_tuning}
\end{figure}

To quantify the extent of this vulnerability, let us consider a standard cosmological setup where the peak threshold ratio is evaluated at a fiducial value of $\nu \equiv \delta_c / \sigma(M) = 8.0$. Utilizing the standard tail-integration formalism, this yields an initial mass fraction of:
\begin{equation}
    \beta_{\text{fid}} \approx \frac{1}{\sqrt{2\pi}\nu}\exp\left(-\frac{\nu^2}{2}\right) \approx 6.22 \times 10^{-16},
\end{equation}
which is a representative order of magnitude required for PBHs to constitute the entirety of the dark matter in the asteroid-mass window. As can be seen in Fig.~\eqref{fig:pbh_fine_tuning}, if a minor modification in the choice of window function or local spatial profile profile increases the effective threshold $\delta_c$ by a mere $10\%$ (shifting $\nu$ from $8.0$ to $8.8$), the exponential suppression forces $\beta(M)$ to plummet to $\approx 6.84 \times 10^{-19}$ --- a reduction by a factor of nearly $910$. Conversely, if $\delta_c$ is lowered by $10\%$ ($\nu = 7.2$), $\beta(M)$ surges up to $\approx 3.01 \times 10^{-13}$, an inflation of nearly three orders of magnitude ($\sim 484$ times larger)~\cite{Mishra:2019pzq}. This dramatic sensitivity demonstrates why the standard analytical evaluation of PBH number density demands an extraordinary level of fine-tuning and why minor structural revisions to the collapse threshold severely impact the final cosmological viability of a given model.

\subsubsection{The Excursion Set and the ``Fudge Factor"}
In the literature, Eq.~\eqref{beta-ps} is frequently multiplied by an empirical factor of 2 (e.g., Refs.~\cite{Kawasaki:2012wr,Garcia-Bellido:2017mdw,Inomata:2018cht,Braglia:2020eai, Palma:2020ejf, Auclair:2026tfy, Iizuka:2026xfc}). This factor originates from the direct application of the Press-Schechter formalism developed for \ac{DM} halo formation. In standard structure formation, the ``cloud-in-cloud" problem arises when small underdense regions are embedded in larger overdense regions, or small overdensities are subsumed by larger ones, leading to a miscounting of collapsed halos. Within the excursion-set framework~\cite{Bond:1990iw} --- where the smoothed density contrast undergoes a stochastic random walk as the smoothing scale varies --- the halo formation process is fundamentally Markovian, mathematically requiring the factor of 2 to resolve the miscounting.

However, recent analysis of the PBH excursion-set dynamics~\cite{Kushwaha:2025zpz} demonstrates that PBH formation is a \emph{non-Markovian} process. Consequently, the standard multiplication by this factor of 2 is irrelevant and should not be applied if the PBHs are forming at horizon-crossing during the \ac{RD} epoch. Conversely, if PBHs form on sub-horizon scales during an \ac{EMD} epoch, the factor of 2 becomes relevant again~\cite{Auclair:2020csm}.

\subsubsection{Physical Implication: Exponential Sensitivity}
Eq.~\eqref{eq:beta_approx} reveals why PBH formation is highly sensitive to the physics of the early universe. The abundance depends exponentially on the ratio $\delta_c / \sigma$ and two specific events play a crucial role: First, during the QCD phase transition ($t \sim 10^{-5}$ s), the equation of state softens ($w < 1/3$), lowering the pressure support. This reduces the collapse threshold $\delta_c$, leading to an exponential enhancement in $\beta(M)$ at the solar mass scale.
Second, during inflation, a small $\mathcal{O}(1)$ increase in the power spectrum amplitude can boost the PBH abundance by many orders of magnitude.

\subsection{From Formation to Today: 
\texorpdfstring{$\Omega_{\text{PBH}}$}{\Omega_{\text{PBH}}}}
It is crucial to explicitly distinguish between the epoch of PBH formation and the present day. The mass fraction $\beta(M)$ refers to the \emph{initial} PBH abundance at formation, encompassing all black holes regardless of their lifetime. Conversely, observational constraints target the present-day density parameter, $\Omega_{\text{PBH}} \equiv \rho_{\text{PBH}}(t_0)/ \rho_{\text{crit}}$, which only includes macroscopic, non-evaporated PBHs ($M \gtrsim 10^{15}$ g).

Because \acp{BH} behave as non-relativistic matter ($\rho \propto a^{-3}$) while the background universe is \ac{RD} ($\rho \propto a^{-4}$), the relative density of PBHs grows linearly with the scale factor $a$ until matter-radiation equality. For non-evaporating PBHs, the relation between the initial fraction and the current density is~\cite{Carr:2009jm}:
\begin{eqnarray}
\Omega_{\text{PBH}} &\approx& 2.7 \times 10^8 \left( \frac{\gamma}{0.2} \right)^{1/2} \nonumber \\ 
&\times& \left( \frac{g_*(t_f)}{106.75} \right)^{-1/3} \left( \frac{M}{M_{\odot}} \right)^{-1/2} \beta(M) \, .   
\end{eqnarray}
Here, $\gamma$ historically represents an average collapse efficiency factor ($\gamma \sim 0.2$), assuming the PBH mass is roughly 20\% of the horizon mass. However, rigorous \ac{NR} demonstrates that PBH formation exhibits near-critical behavior~\cite{Niemeyer:1997mt}. The final mass is not a fixed fraction, but depends continuously on the initial density perturbation profile via the scaling relation $M = \kappa M_H (\delta - \delta_c)^{\upsilon_{\rm c}}$. Here, $\kappa$ is a shape-dependent constant, $M_H$ is the horizon mass, and $\upsilon_{\rm c} \simeq 0.36$ is the universal critical exponent for a radiation fluid~\cite{Musco:2008hv, PhysRevD.103.063538}. $g_*(t_f)$ is the number of relativistic degrees of freedom at formation ($\sim 106.75$ for the Standard Model plateau, dropping to $\sim 10$ after the QCD transition).

The above expression highlights that a microscopically tiny initial fraction (e.g., $\beta \sim 10^{-9}$ for a solar mass PBH) is sufficient to constitute the entirety of the \ac{DM} ($\Omega_{\text{PBH}} \approx 0.26$) today.

Equivalently, one can express this present-day abundance in terms of the comoving number density of PBHs. Because the ratio of the PBH number density to the entropy density, $n_{\text{PBH}}/s$, remains a conserved quantity under adiabatic expansion following formation, one can utilize the relation $\beta(M) \propto M (n/s) T^{-1}$ and the \ac{RD} scaling $M \propto T^{-2}$. As established by Carr \textit{et al.}~\cite{Carr:2021bzv}, the current number density per Gpc$^3$ is cleanly expressed as:
\begin{eqnarray}
n_{\text{PBH}}(t_0) &\approx& 1.25 \times 10^{28} \gamma^{1/2} \left( \frac{g_*(t_f)}{106.75} \right)^{-1/4} \nonumber \\ 
&\times& \left( \frac{M_\odot}{M} \right)^{3/2} \beta(M) \, .
\end{eqnarray}
This exact thermodynamic derivation perfectly recovers the $M^{-3/2}$ scaling anticipated by the geometric site-counting arguments discussed earlier in Sec.~\eqref{sec:Thermo}.

\section{PBH formation in Early Matter-Dominated epoch}
\label{sec:MatterDomination}

The entire paradigm of the \emph{race against sound} and the substantial compaction threshold ($\mathcal{C}_c \approx 0.45$) is predicated on the immense radiation pressure of the early universe ($w = 1/3$). As derived heuristically using the Spherical Top-Hat collapse model (detailed in Appendix (\ref{sec:TopHatDerivation})), the analytic threshold for collapse scales linearly with the equation of state: $\delta_{\text{th}} \approx w$. \emph{Extrapolating this relation} raises an important cosmological question: What happens if a population of PBHs forms during an \ac{EMD} epoch? 

Such epochs are frequently predicted in beyond-Standard-Model cosmology, occurring during a prolonged reheating phase after inflation, or during the oscillation of a heavy scalar field before they decay into radiation~\cite{Assadullahi:2009jc}. During an EMD epoch, the macroscopic equation of state and the sound speed drop essentially to zero ($w \approx 0$, $c_s \approx 0$). Under the strict \textit{perfect fluid approximation}, one might \emph{naively conclude} that the physical barriers to PBH formation completely vanish:
\begin{enumerate}
    \item \textbf{No Jeans Barrier:} Because $c_s \approx 0$, the Jeans length collapses. Even sub-horizon perturbations, which would normally oscillate as acoustic waves in the radiation plasma, are completely free to undergo gravitational collapse.
    \item \textbf{Vanishing Threshold:} The idealized mathematical threshold for the compaction function plummets to $\mathcal{C}_c \to 0$~\cite{Khlopov:1980mg}. 
\end{enumerate}

However, it is crucial to recognize that this vanishing threshold is an artifact of assuming an ideal, spherically symmetric perfect fluid. While it is true that macroscopic isotropic pressure no longer prevents collapse, it is physically incorrect to conclude that PBH formation becomes trivial. When $w \to 0$, the collapsing region loses the \emph{stiffening protection} of radiation pressure and becomes entirely dominated by secondary physical effects that were previously negligible.

Because the collapse is no longer resisted by isotropic pressure gradients, it becomes highly susceptible to initial anisotropies. Deviations from perfect sphericity grow non-linearly, causing the collapsing patch to flatten into a \emph{pancake} or a spindle rather than a \ac{BH}~\cite{Harada:2016mhb}. Furthermore, the conservation of angular momentum implies that even minuscule initial rotations (tidal torques from neighboring perturbations) will halt the radial collapse, leading instead to the formation of a supported accretion disk or a \ac{DM} halo~\cite{Kokubu:2018fxy}. 

Therefore, while the idealized \emph{density threshold} for PBH formation formally approaches zero in a perfect dust fluid, the actual PBH abundance in an EMD epoch is heavily suppressed by strict requirements on spherical symmetry and low initial spin, fundamentally altering the standard mass-scaling predictions. For details, see Refs.~\cite{Assadullahi:2009jc,Kokubu:2018fxy}.

\section{Limitations of the Asymptotically Flat Evaporation Framework}
\label{sec:HawkingRad}

The observational constraints on light PBHs ($M < 10^{15}$ g) are overwhelmingly derived from the assumption of complete evaporation by the present epoch. The original calculation of Hawking radiation, as detailed in~\cite{Hawking:1975vcx}, utilizes a Bogolyubov transformation between vacuum states defined at past and future null infinity ($\mathcal{I}^-$ and $\mathcal{I}^+$). However, this framework is predicated on three key assumptions that do not strictly hold for PBHs formed in the early Universe:

\begin{description}
\item[Asymptotic Non-flatness and choice of vacuum]
The standard derivation assumes that the spacetime is asymptotically flat, allowing for a unique definition of positive-frequency modes at $\mathcal{I}^\pm$ that vary as $f_\omega \propto e^{-i\omega v}$ and $F_\omega \propto e^{-i\omega u}$. However, as discussed in Sec.~\eqref{sec:CosmoBH}, PBHs are nested within a dynamic FLRW background. In such a cosmological context, the metric does not approach Minkowski space at infinity; it approaches the FLRW background where $R_{\mu\nu} \neq 0$. Consequently, there is no global timelike Killing vector to uniquely define a vacuum state or a stationary frequency spectrum, making the \emph{definition of particles} fundamentally observer-dependent and ambiguous \cite{Ashtekar:2004cn}.

\item[The Teleological Nature of the Event Horizon]
Hawking's calculation relies on the formation of an event horizon, a teleological concept requiring knowledge of the entire future history of the spacetime to define the trapped region. For a PBH evolving in a \ac{RD} environment with significant accretion and a time-dependent mass $M(t)$, the event horizon is an impractical diagnostic. Instead, the physics of evaporation must be coupled to the \emph{Apparent Horizon} ($R_{AH}$), a quasi-local concept defined by marginally trapped surfaces. Unlike the static Schwarzschild horizon, the apparent horizon can evolve discontinuously and is subject to local energy-momentum flows, necessitating a dynamic treatment of the evaporation rate.

\item[Coupling to the Primordial Plasma]
Standard Hawking radiation assumes a black hole radiating into an empty vacuum. PBHs, however, are born in a dense fluid where the Hawking temperature $T_{BH} \propto M^{-1}$ competes directly with the background CMBR temperature \cite{Carr:2026hot}. For PBHs heavier than approximately $10^{23}$ g (asteroid-mass), the black hole is cooler than the surrounding plasma and will net-accrete rather than evaporate until the background temperature drops below $T_{BH}$ \cite{Carr:2026hot}. Furthermore, the violent formation process and the lack of stationarity during the early \ac{RD} era mean the \emph{No-Hair theorem} is not strictly valid~\cite{Xavier:2020ulw,Sunny:2023dgz}.
\end{description}

These assumptions must be critically re-evaluated before definitive exclusions can be made.

\subsection{Asymptotic Non-flatness and Cosmological Embedding}

As demonstrated by Xavier \emph{et al.}~\cite{Xavier:2021chn}, exact time-dependent solutions for evaporating BHs embedded in cosmological backgrounds reveal that the lack of an asymptotically timelike Killing vector alters the mass-loss history. Unlike isolated BHs in Minkowski space, a PBH surrounded by a cosmological mass distribution exhibits a decay rate that is highly mass-dependent; while larger masses may initially decay faster, the decay rate reduces significantly as the mass decreases~\cite{Xavier:2021chn}. This implies that the standard conclusion --- that all PBHs lighter than $10^{15}$ g must have evaporated by the present epoch --- may be invalid in a dynamic FLRW background, necessitating a reconsideration of the entire low-mass exclusion window.

\subsection{Strong Gravity and High-Curvature Corrections}

As discussed above, the standard Hawking radiation is derived using the semiclassical approximation on a fixed background. For low-mass PBHs, particularly those approaching the asteroid-mass window ($10^{17}$--$10^{23}$ g), the curvature at the horizon is enormous. The Kretschmann scalar evaluated at the event horizon scales as $K(r_h) \propto M^{-4}$, reaching values of $\mathcal{O}(10^{51})$ m$^{-4}$ for microscopic holes. 

In this extreme curvature regime, the standard Einstein-Hilbert action is strictly insufficient, and higher-order curvature corrections (such as $\mathcal{O}(R^2)$ or Gauss-Bonnet terms) become non-negligible. Theoretical analysis of dimensional reduction from the 4-D Horndeski action to 2-D Callan-Giddings-Harvey-Strominger (CGHS) models indicates that higher-derivative corrections produce appreciable modifications to the trace of the vacuum expectation value of the stress-energy tensor, $\langle T(f) \rangle$~\cite{Mandal:2023kpu}. Because current observational constraints (e.g., the INTEGRAL 511 keV line limits) assume that the Hawking flux for asteroid-mass PBHs is strictly identical to the uncorrected spectrum of larger, stellar-mass black holes, these limits must be treated with caution, as they neglect non-perturbative, high-energy corrections to the emission spectrum.

\subsection{Singularity Resolution and Stable Remnants}

The semiclassical picture of runaway evaporation leading to a physical singularity is generally understood to be a pathology of pushing low-energy effective theories beyond their regime of validity. Incorporating UV-complete descriptions into the gravitational action provides a mechanism for regularization. For example, exact solutions utilizing a phantom Dirac-Born-Infeld (DBI) scalar field demonstrate a mechanism of kinetic regularization. In these models, a dynamical "kinetic stiffness" arises --- analogous to shear thickening in non-Newtonian fluids --- which halts gravitational collapse and excises the central singularity in favor of a regular 2-sphere core~\cite{Parvez:2025wtq}.

Crucially, the thermodynamics of these regular black holes dictate that they do not evaporate completely. Instead, evaporation ceases when the horizon shrinks to meet the regular core, leaving behind an extremal, non-singular relic with a mass roughly at the Planck scale ($\sim 10^{-5}$ g)~\cite{Parvez:2025wtq}. The existence of stable relics resurrects a vast region of parameter space: PBHs with arbitrarily small initial masses could have formed in the early Universe, evaporated down to these extremal states, and now constitute the totality of the \ac{CDM} while entirely evading gamma-ray and CMBR evaporation constraints~\cite{Dvali:2020oqi}.

\section{Conclusions and Future outlook}
\label{sec:Conclusion}

While the broader literature frequently emphasizes phenomenological bounds on the \ac{PBH} dark matter fraction --- ranging from microlensing and CMBR distortions to dynamical friction~\cite{Carr:2020xqk,Carr:2021bzv,Green:2020jor,Escriva:2022duf,Carr:2020gox} --- this review has deliberately centered on the underlying physical machinery. We have systematically reconstructed the theoretical lifecycle of these compact objects: beginning with the non-linear collapse of horizon-sized perturbations, analyzing the ensuing hydrodynamic \emph{race against sound} in the primordial plasma, and culminating in the rigorous quantum mechanics governing their evaporation and potential long-term stability.

Having consolidated this theoretical bedrock, we are now positioned to contextualize these fundamental formation and evolution mechanisms within the rapidly advancing multimessenger observational landscape. The research on PBH has evolved from a theoretical conjecture into a definitive empirical pursuit, and the synergy between GW interferometry, high-redshift galaxy surveys, and refined thermodynamic modeling offers a clear roadmap toward confirming or infirming their existence.

\subsection{Probing the Open Windows}

Identifying PBHs as a primary or partial constituent of \ac{DM} requires penetrating the remaining observational \emph{open windows} and distinguishing them from dynamically assembled stellar populations:
\begin{itemize}
    \item \textbf{The Asteroid-Mass Window ($10^{17}$--$10^{23}$ g):} This remains the primary unconstrained parameter space where PBHs could constitute the entirety of the \ac{DM} ($f_{\text{PBH}} = 1$). Future detection relies on the microlensing of X-ray pulsars to bypass wave-optics limits~\cite{Bai:2018bej}, as well as observing the anomalous depletion of high-mass main-sequence stars in ultra-faint dwarf galaxies due to PBH capture and subsequent stellar destruction~\cite{Esser:2025pnt}.
    \item \textbf{The LVK Mass Range ($\sim 10$--$100 \, M_\odot$):} Confirmation of a primordial subpopulation in current GW catalogs hinges on detecting events in the pulsational pair-instability \emph{upper mass gap} (e.g., GW190521) and observing the characteristically low effective inspiral spins ($\chi_{\text{eff}} \approx 0$) predicted by standard PBH formation statistics~\cite{Clesse:2020ghq,Heydenreich:2024nlk,Belotsky:2018wph}.
    
  \item \textbf{Clustering Signatures and Initial Conditions:} While early literature suggested PBHs might generically form in dense clusters, rigorous statistical analyses have since demonstrated that PBHs born from standard Gaussian curvature perturbations exhibit a nearly Poissonian 
  distribution, with negligible initial clustering~\cite{Ali-Haimoud:2018dau, Desjacques:2018wuu, Auclair:2024jwj}. Consequently, any observation of significant initial spatial correlation --- potentially manifesting as coherence between the source-subtracted cosmic infrared and X-ray backgrounds, or as distinctive caustics in microlensing light curves~\cite{Kashlinsky:2016sdv} --- would not merely confirm the PBH hypothesis, but would act as a direct probe of the early universe. Such clustering would strongly indicate the presence of primordial non-Gaussianity or formation via topological defects~\cite{Franciolini:2021tla}, making spatial distribution a powerful diagnostic tool for the inflationary dynamics that seeded the black holes.  
\end{itemize}

\subsection{Next-Generation Experimental Frontiers}

The next decade will provide orders-of-magnitude increases in sensitivity across both the electromagnetic and gravitational spectra:

\begin{description}
\item[\textbf{mHz - KHz GW detectors}] Third-generation (3G) detectors, such as the CE and ET, will probe binary mergers at redshifts $z > 30$, an epoch preceding star formation where stellar remnant black holes cannot exist~\cite{Punturo:2010zz, LIGOScientific:2021job, ET:2019dnz}. Simultaneously, space-borne missions like LISA, Taiji, and TianQin will access the mHz band to probe planetary-mass PBHs and the stochastic background of \ac{SIGWs}~\cite{TianQin:2015yph,Hu:2017mde,LISA:2022yao}.
\item[\textbf{Radio Astronomy}] The SKA will meticulously map the reionization history of the universe. Accreting or evaporating PBHs would inject energy into the intergalactic medium, altering the 21-cm absorption signal and providing a definitive diagnostic of PBH abundance during the Dark Ages~\cite{Camera:2014bwa,Mittal:2021egv}.
\item[\textbf{High-Energy Astrophysics}] Observatories like High-Altitude Water Cherenkov Observatory (HAWC) and the Cherenkov Telescope Array (CTA) will search for the terminal stages of Hawking evaporation, potentially revealing characteristic high-energy gamma-ray bursts from exploding PBHs~\cite{HAWC:2014ycj}.
\item[\textbf{High-Frequency GWs (MHz-GHz)}] Beyond the standard interferometric bands, the frontier of high-frequency gravitational waves offers a direct probe into the mergers of ultra-light, planetary-mass PBHs~\cite{Aggarwal:2025noe}. Recently, a bulk acoustic wave (BAW) detector experiment operating in the MHz regime reported the detection of two rare transient events during 153 days of operation~\cite{Goryachev:2021zzn}. If these signals are confirmed to be of genuine astrophysical origin, the MHz frequencies would perfectly align with the inspiral and merger of PBHs in the $M \sim 10^{27}$ g mass range. Developing this acoustic detector technology~\cite{Campbell:2025mks} could open an entirely new window into the previously inaccessible asteroid- and planetary-mass parameter space.
\end{description}

\subsection{Theoretical Pitfalls}

Crucially, as we push toward precision cosmology, the theoretical caveats highlighted in this review must be rigorously addressed. Current constraints on PBH abundance in the low-mass regime often rely on idealized semiclassical approximations that may fail in realistic contexts. As detailed in Section~(\ref{sec:HawkingRad}), the breakdown of asymptotic flatness in dynamic FLRW backgrounds~\cite{Xavier:2021chn, Ashtekar:2004cn}, the potential for quantum backreaction via the "Memory Burden"~\cite{Dvali:2018xpy, Alexandre:2024nuo, Thoss:2024hsr}, the modification of Hawking flux by higher-curvature corrections~\cite{Mandal:2023kpu}, and the resolution of singularities into stable, Planck-mass relics~\cite{Parvez:2025wtq} all point toward a common conclusion: light PBHs may not evaporate as cleanly or completely as standard Hawking theory predicts. These non-perturbative and cosmological effects necessitate a fundamental re-evaluation of current gamma-ray and CMBR exclusion bounds.

The standard theoretical frameworks for both Hawking evaporation and \ac{GW} generation heavily rely on the assumption of asymptotically flat spacetimes. However, PBHs are inextricably embedded within a dynamic, expanding \ac{FLRW} background. For quantum evolution, the lack of a global timelike Killing vector fundamentally alters the mass-loss history and makes the definition of asymptotic particle states observer-dependent~\cite{Xavier:2021chn, Ashtekar:2004cn}. Furthermore, this assumption directly impacts current observational inference. The numerical relativity templates employed by the \ac{LVK} collaboration to extract binary parameters (e.g., mass, spin, and eccentricity) are constructed using asymptotically flat boundary conditions, with cosmic expansion treated merely as a kinematic redshift scaling. Neglecting non-linear cosmological tail effects and the true dynamic embedding of the binary can introduce systematic biases in parameter estimation~\cite{Ashtekar:2015ooa}. Consequently, inferring a purely primordial origin from current GW catalogs requires caution, as the underlying waveform templates do not yet fully encapsulate the cosmological nature of the spacetime.

Furthermore, the precise mathematical definition of the \ac{PBH} boundary in a dynamic \ac{FLRW} background introduces significant theoretical subtleties. Because the global event horizon is teleological and uncomputable in an evolving universe, numerical analyses must rely purely on local optical scalars to identify the apparent horizon---specifically locating marginally trapped surfaces where the expansion of outgoing null congruences vanishes ($\theta = 0$). While classical collapse guarantees a central singularity via the Raychaudhuri equation and the Penrose singularity theorems~\cite{Penrose:1964wq,Landsman:2022hrn}, applying this strict geometric machinery to the ultimate \ac{PBH} remnant is problematic. If extreme-curvature quantum effects or backreaction violate the underlying energy conditions, the classical convergence of null congruences predicted by the Raychaudhuri equation may fail, potentially replacing the singularity with a regular, non-singular core~\cite{Parvez:2025wtq}.

\subsection{The Ultimate Cosmological Probe}

While the current catalog of GW transients has revolutionized our understanding of black hole astrophysics, distinguishing a PBH from a dynamically assembled stellar black hole in the $\mathcal{O}(10) \, M_{\odot}$ range remains a profound statistical challenge. However, the detection of a \emph{sub-solar} mass black hole ($M \ll 1 \, M_{\odot}$) would instantly break this degeneracy. Because the remnants of standard stellar evolution are strictly bounded from below by the Chandrasekhar and Tolman-Oppenheimer-Volkoff limits, no known astrophysical mechanism can produce a macroscopic black hole lighter than a neutron star. 

Consequently, the observation of a sub-solar mass inspiral chirp by current ground-based interferometers, or by next-generation observatories, would provide absolute, \emph{smoking-gun} evidence for the PBH paradigm~\cite{LIGOScientific:2021job, ET:2019dnz}. Such a detection would not only solve the \ac{DM} anomaly but would effectively transform black holes into the ultimate cosmological probes---allowing us to read the precise hydrodynamic and quantum initial conditions of the universe mere fractions of a second after the Big Bang.

While the focus of this review has remained on the formation and distribution of PBHs, recent advancements have highlighted their potential as unique probes of the high-curvature regime through indirect observational signatures. A promising avenue involves the interaction of asteroid-mass PBHs with interstellar media, specifically within the neutral hydrogen ($\text{H}_{\text{I}}$) clouds ubiquitous throughout the galaxy. As demonstrated recently~\cite{Christopher:2026mvm}, the intense local gravitational curvature of a sub-lunar PBH acts as a microscopic tidal perturbator, inducing a quantum-mechanical splitting of the bound internal energy levels of neutral hydrogen atoms. This perturbation fundamentally alters the standard 9.9 GHz absorption profile of interstellar hydrogen, redistributing the single spectral line into a distinct, high-bandwidth cluster of thousands of shifted absorption features spanning approximately 2 GHz. This phenomenon --- termed as \emph{Gravitational Spectral Radio Forest} --- suggests that neutral hydrogen gas effectively functions as a high-precision quantum sensor for gravitational fields. Consequently, such wideband chaotic radio signatures provide a new, distinct observational channel to constrain the abundance of asteroid-mass PBHs, complementing traditional multi-messenger search strategies.

Ultimately, the PBH paradigm confronts the central pathology of classical GR: the spacetime singularity. The Penrose-Hawking theorems dictate that classical gravitational collapse inevitably terminates in a singularity, marking a fundamental breakdown of predictability. However, the formation, thermodynamic evolution, and structural stability of PBHs --- particularly low-mass relics governed by extreme-curvature dynamics ---strictly require a UV-complete mechanism that halts collapse and regularizes the core. Consequently, the definitive detection of a PBH by future observatories would carry an implication far beyond astrophysics. It would provide direct observational evidence that classical singularities do not exist in nature. Beyond resolving the \ac{DM} anomaly, the discovery of a PBH would serve as the first empirical confirmation of singularity resolution, transforming these primordial relics into the ultimate, observable testbed for quantum gravity.

\vspace{0.5cm}
\noindent\emph{\underline{Acknowledgement:}} The authors are grateful to P.~Bansal, I.~Chakraborty, S.~Mahesh Chandran, A.~Choudhury, P. G. Christopher, Joseph~P.~J, K.~Hari, A. Kushwaha, S.~Mandal, A.~Naskar and T.~Parvez for their valuable discussions and feedback on the earlier draft. A. V. is supported by the Council of Scientific \& Industrial Research (CSIR), India under its Research Associateship program. The work is supported by ANRF-Advanced Research Grant (ANRF/ARG/2025/001514/PS).  

\appendix 
\section{Jeans Instability: Race against time}
\label{app:Jeans-RaceAgainstTime}

To have a physical understanding of this process, let us assume the overdense region to be a sphere of gas that has just stopped expanding (at the turnaround point) and is trying to collapse under its own gravity. At this point in time, there are two competing processes:
\begin{enumerate}
\item \textbf{Gravity (The Collapse):} Gravity tries to pull the outer shell of the sphere inward to the center to form a \ac{BH}.
\item \textbf{Pressure (The Escape):} The high density creates high pressure. A \emph{pressure wave} (sound wave) starts at the center and travels outward, trying to push the gas apart and smooth out the density.
\end{enumerate}

We can view this as a race against time.
\begin{enumerate}
\item \textbf{Time to Collapse ($t_{\text{coll}}$):} This is roughly the ``free-fall time" --- how long it takes for gravity to crush the sphere to a point.

\item \textbf{Time for Sound to Cross ($t_{\text{cross}}$):} This is how long it takes for a pressure wave to travel from the center to the edge of the sphere ($R$). Since pressure travels at the speed of sound ($c_s$), this time is $t_{\text{cross}} \approx R / c_s$.
\end{enumerate}

In order for the \ac{BH} to be formed, the sphere must collapse \emph{before} the pressure wave can cross the region and tell the outer layers to stop falling. Mathematically, this means:
\begin{equation}
t_{\text{coll}} < t_{\text{cross}}
\end{equation}
Let us now go back to the top-hat model we discussed earlier and obtain the relevant variables:
\begin{enumerate}
\item \textbf{The Size of the Region ($R$):} 
At turnaround, the size of the region is $R \approx b_{max}$.
From the derivation in Appendix \eqref{sec:TopHatDerivation}, we see that $b_{max} \propto \delta^{-1/2}$. This physically means that a \emph{larger initial density contrast} $\delta$ leads to a \emph{smaller}, more compact universe at turnaround. 

\item \textbf{The Time Scale ($t$):} The time available for this process is the age of the universe at that moment, $t_{max}$.
For the spherical top hat we have, $t_{max} \approx t_i \delta^{-1}$.
This physically means that for a larger $\delta$ the collapse happens much faster (earlier).
\end{enumerate}

We now substitute these into the condition 
\begin{equation}
    t_{\text{coll}} < t_{\text{cross}} \quad 
\Longrightarrow \quad t_{\text{coll}} < \frac{R}{c_s}
\end{equation}
Rearranging for $R$:
\begin{equation}
R > c_s \times t_{\text{coll}}
\end{equation}
Comparing the above expression with Eq.~\eqref{eq:JeansCondition}, we see that they are identical. In other words the above condition is  the \emph{Jeans Criterion}, i. e., \emph{the physical size ($R$) must be larger than the distance sound can travel ($c_s t$)}.

We now rewrite the above expression interms of $\delta$. Substitute the above scalings we get:
\begin{equation}
R \propto \delta^{-1/2} \quad t \propto \delta^{-1} \, .
\end{equation}
This leads to the condition: 
\begin{equation}
\delta \equiv \frac{\delta \rho}{\overline{\rho}} > \frac{c_s^2}{c^2}
\end{equation}
The above expression provides a nice way of understanding PBH formation (see also Fig. \eqref{fig:PBHformation}): 
\begin{enumerate}
\item  \textbf{If $\delta$ is small:} The overdensity is physically spread out! Hence, it takes a long time to turn around. This gives the pressure waves plenty of time to cross the region, smooth out the density, and prevent a \ac{BH}.

\item \textbf{If $\delta$ is large ($\delta > 1/3$):} The overdensity is compact and has very strong gravity. It turns around and collapses so quickly that the pressure waves (moving at $c_s$) are too slow to cross the region before it's too late. Gravity wins, and a PBH forms.
\end{enumerate}
Thus, the speed of sound ($c_s$) sets the scale. If the speed of sound is high (harder to compress), a higher density $\delta$ is required for the PBH formation. Since, in the \ac{RD} era, sound is very fast ($c_s \approx 0.58c$), a large density contrast ($\delta > 1/3$) is required for PBH formation.

\section{Detailed Derivation of Newtonian Jeans Instability}
\label{app:JeansDetail}

In this Appendix, we provide a detailed derivation of the background evolution and the linear perturbation equations for a self-gravitating fluid in Newtonian gravity, leading to the Jeans instability criterion.

We consider a non-relativistic fluid described by mass density $\rho(\mathbf{X}, t)$, pressure $p(\mathbf{X}, t)$, velocity field $\mathbf{v}(\mathbf{X}, t)$, and gravitational potential $\phi(\mathbf{X}, t)$. The governing equations are:
\begin{align}
\label{eq:cont}
\text{Continuity:} & \quad \frac{\partial \rho}{\partial t} + \nabla \cdot (\rho \mathbf{v}) = 0 \, , \\
\label{eq:euler}
\text{Euler:} & \quad \frac{\partial \mathbf{v}}{\partial t} + (\mathbf{v} \cdot \nabla) \mathbf{v} = -\frac{\nabla p}{\rho} - \nabla \phi \, , \\
\label{eq:poisson}
\text{Poisson:} & \quad \nabla^{2} \phi = 4\pi G \rho \, .
\end{align}
where $\nabla$ denotes the derivative with respect to the physical Euclidean coordinate $\mathbf{X}$. Note that we use the reduced Planck mass relation $4\pi G = 1/(2M_{\mathrm{Pl}}^2)$.

\subsection{Background Solution}

We first define a homogeneous and isotropic background where the background density $\bar{\rho}$ depends only on time, and the velocity follows the Hubble flow:
\begin{equation}
\bar{\rho} = \bar{\rho}(t), \quad \bar{p} = \bar{p}(t), \quad \bar{\mathbf{v}} = H(t) \mathbf{X} \, .
\end{equation}
where $H(t) \equiv \dot{a}/a$ is the Hubble parameter and $a(t)$ is the scale factor. Substituting these into Eq.~\eqref{eq:cont}, we get:
\begin{equation}
\dot{\bar{\rho}} + \bar{\rho} H (\nabla \cdot \mathbf{X}) = \dot{\bar{\rho}} + 3H \bar{\rho} = 0 \, .
\end{equation}
This leads to:
\begin{equation}
\bar{\rho}(t) = \rho_{0} \left( \frac{a_{0}}{a(t)} \right)^3 \, .
\end{equation}
corresponding to the matter dominated epoch. It is consistent with the Newtonian approximation we started with.

Now, substituting $\bar{\rho}$ into Eq.~\eqref{eq:poisson}:
\begin{equation}
\nabla^2 \bar{\phi} = 4 \pi G {\bar{\rho}} \, .
\end{equation}
Since the RHS is a function of time only, the solution of the above Laplace equation is analogous to $\nabla^2 f = C$ where $C$ is a constant. As we know it is quadratic in $\mathbf{X}$. Thus:
\begin{equation}
\bar{\phi} = \frac{2 \pi G}{3} \bar{\rho} |\mathbf{X}|^2 \, .
\end{equation}
Lastly, we will look at the Euler equation, especially the convection term. For the background we have $(\bar{\mathbf{v}} \cdot \nabla) \bar{\mathbf{v}} = (H \mathbf{X} \cdot \nabla) (H \mathbf{X}) = H^2 \mathbf{X}$.
The time derivative is $\partial \bar{\mathbf{v}} / \partial t = \dot{H} \mathbf{X}$.
Substituting into Eq.~\eqref{eq:euler} (neglecting pressure gradients for the homogeneous background):
\begin{equation}
\dot{H} \mathbf{X} + H^2 \mathbf{X} = -\nabla \bar{\phi} = - \frac{4 \pi G\bar{\rho}}{3} \mathbf{X} \, .
\end{equation}
Removing $\mathbf{X}$, we get the acceleration equation:
\begin{equation}
\frac{\ddot{a}}{a} = - \frac{4\pi G\bar{\rho}}{3} \, .
\end{equation}
Substituting $\bar{\rho} = C/a^3$ into the above equation, we get:
\begin{equation}
\ddot{a} = - \frac{4 \pi G C}{3} \frac{1}{a^2} \, .
\end{equation}
Now, we multiply both sides by the integrating factor $2\dot{a}$:
\begin{equation}
2\dot{a}\ddot{a} = - \frac{C}{3 M_{\mathrm{Pl}}^{2}} \frac{\dot{a}}{a^2} \, .
\end{equation}
We can now integrate both sides with respect to time:
\begin{equation}
\int \frac{d}{dt}(\dot{a}^2) dt = - \frac{C}{3 M_{\mathrm{Pl}}^{2}} \int \frac{d}{dt}\left(-\frac{1}{a}\right) dt \, .
\end{equation}
This yields:
\begin{equation}
\dot{a}^2 = \frac{C}{3 M_{\mathrm{Pl}}^{2}} \frac{1}{a} - K \, ,
\end{equation}
where $-K$ is the integration constant. Finally, substituting back $C = \bar{\rho} a^3$ and dividing by $a^2$:
\begin{equation}
\frac{\dot{a}^2}{a^2} = \frac{\bar{\rho} a^3}{3 M_{\mathrm{Pl}}^{2} a^3} - \frac{K}{a^2} \quad \Longrightarrow \quad H^2 = \frac{\bar{\rho}}{3 M_{\mathrm{Pl}}^{2}} - \frac{K}{a^2} \, .
\end{equation}

\subsection{Linear Perturbations}
We introduce small perturbations:
\begin{equation}
\rho = \bar{\rho} + \delta \rho, \quad \mathbf{v} = \bar{\mathbf{v}} + \delta \mathbf{v}, \quad \phi = \bar{\phi} + \delta \phi, \quad p = \bar{p} + \delta p \, .
\label{eq:Pert-Newtonian}
\end{equation}
We substitute these into the fundamental equations and keep only first-order terms. The continuity equation \eqref{eq:cont} becomes 
\begin{equation}
\frac{\partial (\bar{\rho} + \delta \rho)}{\partial t} + \nabla \cdot [(\bar{\rho} + \delta \rho)(\bar{\mathbf{v}} + \delta \mathbf{v})] = 0 \, .
\end{equation}
Using $\nabla \cdot \bar{\mathbf{v}} = 3H$ and subtracting the background equation:
\begin{equation}
\dot{\delta \rho} + 3H \delta \rho + \bar{\mathbf{v}} \cdot \nabla \delta \rho + \bar{\rho} \nabla \cdot \delta \mathbf{v} = 0 \, .
\end{equation}
Since $\bar{\mathbf{v}} = H\mathbf{X}$:
\begin{equation}
\label{eq:pert_cont}
\frac{\partial \delta \rho}{\partial t} + 3H \delta \rho + H \mathbf{X} \cdot \nabla \delta \rho + \bar{\rho} \nabla \cdot \delta \mathbf{v} = 0 \, .
\end{equation}

Substituting Eq.~\eqref{eq:Pert-Newtonian} in the Euler equation \eqref{eq:euler}, we get:
\begin{equation}
\frac{\partial (\bar{\mathbf{v}} + \delta \mathbf{v})}{\partial t} + [(\bar{\mathbf{v}} + \delta \mathbf{v}) \cdot \nabla] (\bar{\mathbf{v}} + \delta \mathbf{v}) = -\frac{\nabla (\bar{p} + \delta p)}{\bar{\rho}} - \nabla (\bar{\phi} + \delta \phi) \, .
\end{equation}
The convective term expands as $(\bar{\mathbf{v}} \cdot \nabla)\bar{\mathbf{v}} + (\bar{\mathbf{v}} \cdot \nabla)\delta \mathbf{v} + (\delta \mathbf{v} \cdot \nabla)\bar{\mathbf{v}}$.
Note that $(\delta \mathbf{v} \cdot \nabla)\bar{\mathbf{v}} = (\delta \mathbf{v} \cdot \nabla)(H\mathbf{X}) = H \delta \mathbf{v}$.
Subtracting the background yields:
\begin{equation}
\label{eq:pert_euler}
\frac{\partial \delta \mathbf{v}}{\partial t} + H \mathbf{X} \cdot \nabla \delta \mathbf{v} + H \delta \mathbf{v} = - \frac{\nabla \delta p}{\bar{\rho}} - \nabla \delta \phi \, .
\end{equation}
Lastly, the Poisson equation becomes:
\begin{equation}
\nabla^2 \delta \phi = \frac{1}{2 M_{\mathrm{Pl}}^{2}} \delta \rho \, .
\end{equation}
We now move to the Fourier domain corresponding to the  comoving coordinates $\mathbf{x} = \mathbf{X}/a$. Specifically, we introduce  $\delta \rho$ in comoving coordinates as:
\begin{equation}
\delta \rho(\mathbf{X}, t) = \int \frac{d^3k}{(2\pi)^3} e^{i \frac{\mathbf{k} \cdot \mathbf{X}}{a(t)}} \delta \rho_{\mathbf{k}}(t) \, .
\end{equation}
Crucially, the time derivative at fixed $\mathbf{X}$ acts on the basis function:
\begin{equation}
\frac{\partial}{\partial t} \left[ e^{i \frac{\mathbf{k} \cdot \mathbf{X}}{a}} \right] = - i \frac{\mathbf{k} \cdot \mathbf{X}}{a} H e^{i \dots} = - (\mathbf{X} \cdot \nabla) H e^{i \dots} \, .
\end{equation}
This term cancels the $H \mathbf{X} \cdot \nabla$ convective terms in Eqs.~\eqref{eq:pert_cont} and \eqref{eq:pert_euler}.

Defining the density contrast $\delta_{\mathbf{k}} = \delta \rho_{\mathbf{k}} / \bar{\rho}$ and assuming adiabatic pressure $\delta p = c_s^2 \delta \rho$, the continuity equation becomes:
\begin{equation}
\dot{\delta}_{\mathbf{k}} + \frac{i \mathbf{k} \cdot \delta \mathbf{v}_{\mathbf{k}}}{a} = 0 \, .
\end{equation}
The perturbed Euler equation becomes 
\begin{equation}
\frac{d}{dt} \left( \frac{i \mathbf{k} \cdot \delta \mathbf{v}_{\mathbf{k}}}{a} \right) + H \left( \frac{i \mathbf{k} \cdot \delta \mathbf{v}_{\mathbf{k}}}{a} \right) = \frac{k^2}{a^2} c_s^2 \delta_{\mathbf{k}} + \frac{k^2}{a^2} \delta \phi_{\mathbf{k}} \, .
\end{equation}
Substituting $\delta \phi_{\mathbf{k}} = -(a^2/k^2)(1/2M_{Pl}^2) \bar{\rho} \delta_{\mathbf{k}}$ from Poisson:
\begin{equation}
\frac{d}{dt} \left( -\dot{\delta}_{\mathbf{k}} \right) + H \left( -\dot{\delta}_{\mathbf{k}} \right) = \left( \frac{k^2 c_s^2}{a^2} - \frac{\bar{\rho}}{2 M_{\mathrm{Pl}}^{2}} \right) \delta_{\mathbf{k}} \, .
\end{equation}
Rearranging gives the final Jeans Stability Equation:
\begin{equation}
\label{eq:Newtonian-delta}
\ddot{\delta}_{\mathbf{k}} + 2H \dot{\delta}_{\mathbf{k}} + \left( \frac{c_s^2 k^2}{a^2} - \frac{\bar{\rho}}{2 M_{\mathrm{Pl}}^{2}} \right) \delta_{\mathbf{k}} = 0 \, .
\end{equation}
We can define a new scale, \emph{Jeans Scale} $k_J$:
\begin{equation}
\frac{k_J}{a} = \frac{\sqrt{4\pi G \bar{\rho}}}{c_s} \sim \frac{H}{c_s} \, .
\end{equation}
This scale shows that the growth ($\delta_{\mathbf{k}}$ increasing) occurs only when the term in the parenthesis is negative, i.e., when gravity beats pressure ($k < k_J$). In other words, for scales below $k_J$, pressure supports the fluid against gravity.

In the \ac{RD} era ($c_s \approx 1/\sqrt{3}$), the Jeans scale is comparable to the horizon scale. This implies that only perturbations entering the horizon with sufficient amplitude to overcome this pressure barrier can collapse.

\subsection{The Top-Hat Analytical Model}
A simple analytic estimate for the collapse threshold can be derived using the ``Separate Universe" approach, discussed in detailed in Appendix \eqref{sec:GradientExpansion}. Consider a spherical overdensity (Top-Hat profile) evolving as a closed FLRW universe ($K =+1$) embedded in a flat background ($K =0$).

The Friedmann equation for the overdense region (scale factor $b$) and background (scale factor $a$) yields a relation between the density contrast $\delta$ and the curvature perturbation. See the derivation in the following Appendix. Collapse begins when the overdense region stops expanding ($\dot{b}=0$).
Using the condition that the size of the region must exceed the Jeans length at the moment of turnaround ($b_c R > c_s H_c^{-1}$), one obtains the famous analytic threshold approximation \cite{Carr:1975qj}:
\begin{equation}
\delta_{\mathrm{th}} \approx w = \frac{1}{3} \, .
\end{equation}
While this simple $\delta_{\text{th}} \approx 1/3$ result is historically significant, modern \ac{NR} (discussed in the main text) reveals that the threshold is profile-dependent and typically higher ($\delta_c \sim 0.45$).

\section{The Gradient Expansion Approximation}
\label{sec:GradientExpansion}

To study the formation of PBHs rigorously, we must define the initial conditions deep in the \ac{RD} era, long before the \ac{BH} horizon forms. At this epoch, the scale of the collapsing perturbation $L$ is much larger than the cosmological horizon $R_H \sim 1/H$ (i.e., it is super-horizon).

Standard linear perturbation theory breaks down because PBH formation requires large, non-linear density contrasts ($\delta \gtrsim 1$). Instead, we employ the \emph{Gradient Expansion} (or long-wavelength) approximation \cite{Tanaka:2006zp,Polnarev:2006aa}.

\subsection{Physical Intuition: The Separate Universe Approach}
\label{sec:SeparateUniverse}

The mathematical validity of the gradient expansion rests on the \emph{Separate Universe Assumption}~\cite{Wands:2000dp}. 

On super-horizon scales ($k \ll aH$), spatial gradients are suppressed by the expansion parameter $\epsilon$. Physically, this means that causal physics (like pressure waves) cannot propagate between distant regions within a Hubble time. Consequently, we can model the inhomogeneous universe as a patchwork of locally homogeneous, independent fluid elements. Each super-horizon patch evolves like a separate \ac{FLRW} universe with its own local scale factor $\tilde{a}(t, \mathbf{x})$:
\begin{equation}
\tilde{a}(t, \mathbf{x}) = a(t) e^{\zeta(t, \mathbf{x})} \, .
\end{equation}
Here, the curvature perturbation $\zeta$ acts as a local perturbation to the logarithmic expansion.

\subsubsection*{Adiabatic Perturbations}
\begin{figure}[h!]
\centering
\includegraphics[width=0.4\textwidth]{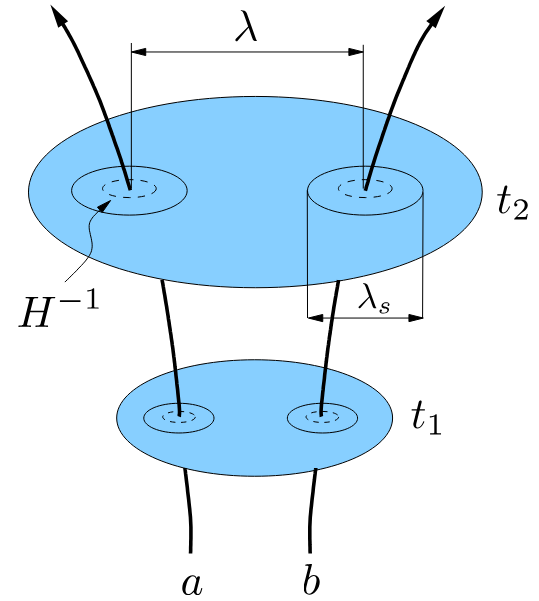} 
\caption{Illustration of the separate universe approach (Reproduced from Wands et al. \cite{Wands:2000dp}). Two super-horizon regions evolve independently. For adiabatic perturbations, they follow the same trajectory in phase space, merely shifted in time, leading to a conserved curvature perturbation $\zeta$.}
\label{fig:separate_universe}
\end{figure}

For adiabatic fluctuations, the matter content in all patches follows the same unique equation of state $P(\rho)$. There are no variations in composition (entropy) between regions. 
As illustrated in Fig.~(\ref{fig:separate_universe}), consider two regions (a) and (b). Since they share the same physical history, their evolution trajectories are identical in phase space. The only difference is where they effectively sit on the time axis. Region (b) can be viewed as Region (a) evolved by a slightly different time interval.

Mathematically, the local density $\rho(t, \mathbf{x})$ can be related to the background density $\bar{\rho}(t)$ by a local time shift $\delta t(\mathbf{x})$:
\begin{equation}
\rho(t, \mathbf{x}) = \bar{\rho}(t + \delta t(\mathbf{x})) \, .
\end{equation}
Because the local physics is identical, the expansion from one uniform-density slice to the next is uniform. Consequently, the "misalignment" between the patches remains constant, implying that the curvature perturbation is conserved:
\begin{equation}
\label{eq:Adiabaticlimit}
\dot{\zeta} = 0 \quad (\text{Adiabatic Limit}) \, .
\end{equation}

\subsubsection*{Non-Adiabatic Perturbations}
For non-adiabatic (isocurvature) perturbations, this simple picture breaks down. If there is a pressure perturbation $\delta P_{\text{nad}}$ independent of density (e.g., due to multiple fields), different patches will have different equations of state. This drives a differential expansion rate between region (a) and (b) that is not merely a time shift. As a result, the curvature perturbation $\zeta$ evolves in time on super-horizon scales.

\subsection{The Expansion Parameter}
The approximation relies on the smallness of spatial gradients compared to time derivatives. We define a small parameter $\epsilon$:
\begin{equation}
\epsilon \equiv \frac{R_H}{L} \sim \frac{k}{aH} \ll 1 \, .
\end{equation}
We expand the Einstein equations in powers of $\epsilon$.
\begin{itemize}
    \item At zeroth order ($\epsilon^0$), the spatial gradients vanish. Each local patch of the universe evolves independently as a separate FLRW universe (the "Separate Universe" approach).
    \item At leading non-trivial order, the spatial curvature becomes non-zero but remains constant in time.
\end{itemize}

\subsubsection{The Non-Linear Metric}
In the synchronous comoving gauge, the general line element can be written using the 3+1 decomposition. Applying the gradient expansion, the metric at leading order takes the \emph{quasi-isotropic} form \cite{Shibata:1999zs}:
\begin{equation}
\label{eq:GradientMetric}
ds^2 = -dt^2 + a^2(t) e^{2\zeta(r)} \left( dr^2 + r^2 d\Omega^2 \right) + \mathcal{O}(\epsilon^2) \, .
\end{equation}
This form is physically significant for three reasons:
\begin{enumerate}
    \item \textbf{Conformal Factor:} The term $e^{2\zeta(r)}$ acts as a local conformal factor. It represents the non-linear curvature perturbation.
    \item \textbf{Conservation of $\zeta$:} A key result of the gradient expansion is that, for adiabatic perturbations, the curvature profile $\zeta(r)$ is a conserved quantity ($\dot{\zeta} = 0$) on super-horizon scales \cite{Lyth:2004gb,Tanaka:2006zp}. This allows us to specify $\zeta(r)$ as a time-independent "initial condition" set by inflation.
    \item \textbf{Asymptotic Behavior:} As $r \to \infty$, we require $\zeta(r) \to 0$, ensuring the metric asymptotically approaches the background FLRW universe.
\end{enumerate}

\subsection{Derivation of Density Contrast}
\label{sec:DerivationDelta}

The relation between the curvature perturbation $\zeta$ and the density contrast $\delta$ arises directly from the Einstein Field Equations, specifically the time-time component (the Hamiltonian Constraint).

In the Arnowitt–Deser–Misner (ADM) formulation (or 3+1 decomposition of General Relativity), the Hamiltonian constraint is given by the Gauss-Codazzi relation projected onto the spatial hypersurface:
\begin{equation}
\label{eq:HamiltonianGeneral}
{}^{(3)}R + K^2 - K_{ij}K^{ij} = 16\pi G \rho \, .
\end{equation}
where, ${}^{(3)}R$ is the Ricci scalar of the 3-dimensional spatial metric $\gamma_{ij}$, $K_{ij}$ is the extrinsic curvature tensor, and $K = \gamma^{ij}K_{ij}$ is its trace and $\rho$ is the total energy density measured by the Eulerian observer.

In the synchronous comoving gauge, the gradient expansion metric (Eq.~(\ref{eq:GradientMetric})) is:
\begin{equation}
\gamma_{ij} = a^2(t) e^{2\zeta(r)} \tilde{\gamma}_{ij} \, ,
\end{equation}
where $\tilde{\gamma}_{ij}$ is the metric of flat space (in spherical coordinates).
The extrinsic curvature is defined as the Lie derivative along the normal vector:
\begin{equation}
K_{ij} = \frac{1}{2} \frac{\partial \gamma_{ij}}{\partial t} =
\frac{1}{2} \left( 2H \gamma_{ij} + 2\dot{\zeta} \gamma_{ij} \right) \, .
\end{equation}
A key property of the gradient expansion on super-horizon scales is that $\zeta$ is conserved in time, so $\dot{\zeta} \approx 0$ (up to $\mathcal{O}(\epsilon^2)$ corrections). Thus:
\begin{equation}
K_{ij} \approx H \gamma_{ij} \, .
\end{equation}
This leads to the extrinsic curvature terms:
\begin{equation}
K = \gamma^{ij} (H \gamma_{ij}) = 3H \, , \quad \quad K_{ij}K^{ij} = (H \gamma_{ij})(H \gamma^{ij}) = 3H^2 \, .
\end{equation}
This leads to:
\begin{equation}
K^2 - K_{ij}K^{ij} = (3H)^2 - 3H^2 = 6H^2 \, .
\end{equation}

Since the 3-Space is a conformally flat metric $\gamma_{ij} = \Psi^4 \delta_{ij}$ (where $\Psi = a^2 e^{2\zeta} \tilde{\gamma}_{ij}$), we can use the conformal relation for $e^{2\zeta}$) leading to:
\begin{equation}
{}^{(3)}R = \frac{1}{a^2 e^{2\zeta}} \left( \tilde{R} - 4\tilde{\nabla}^2\zeta - 2(\tilde{\nabla}\zeta)^2 \right) \, .
\end{equation}
where  $\tilde{\gamma}_{ij}$ is Ricci scalar corresponding to $\gamma_{ij}$ which is zero. Hence, 
\begin{equation}
{}^{(3)}R = -\frac{2}{a^2} e^{-2\zeta} \left( 2\tilde{\nabla}^2\zeta + (\tilde{\nabla}\zeta)^2 \right) \, .
\end{equation}

We now substitute the above expressions for $K$ and ${}^{(3)}R$ back into the Hamiltonian constraint (Eq.~\eqref{eq:HamiltonianGeneral}), leading to:
\begin{equation}
-\frac{2}{a^2} e^{-2\zeta} \left( 2\tilde{\nabla}^2\zeta + (\tilde{\nabla}\zeta)^2 \right) + 6H^2 = 16\pi G \rho \, .
\end{equation}
Important to note that this is a general expression and we have not not made any approximation (except for the fact that $\zeta$ is conserved in the super-Hubble scales).  We now split the density into a background and perturbation: $\rho(t,r) = \overline{\rho}(t) + \delta \rho(t,r)$.
From the background Friedmann equation, we know $6H^2 = 16\pi G \overline{\rho}(t)$. Subtracting this background part leaves the perturbation:
\begin{equation}
-\frac{2}{a^2} e^{-2\zeta} \left( 2\tilde{\nabla}^2\zeta + (\tilde{\nabla}\zeta)^2 \right) = 16\pi G \delta \rho \, .
\end{equation}
Dividing by $16\pi G \overline{\rho}(t) = 6H^2$, we isolate the density contrast:
\begin{equation}
\delta \equiv \frac{\delta \rho}{\overline{\rho}(t)} = -\frac{2}{3} \frac{1}{a^2 H^2} e^{-2\zeta} \left( \tilde{\nabla}^2\zeta + \frac{1}{2}(\tilde{\nabla}\zeta)^2 \right) \, .
\end{equation}

Finally, we expand the Laplacian $\tilde{\nabla}^2$ and gradient $(\tilde{\nabla})^2$ in spherical coordinates:
\begin{equation}
\tilde{\nabla}^2 \zeta = \zeta''(r) + \frac{2}{r}\zeta'(r) \, , \quad \quad (\tilde{\nabla}\zeta)^2 = (\zeta'(r))^2 \, .
\end{equation}
Substituting these yields the explicit radial dependence:
\begin{equation}
\frac{\delta \rho}{\overline{\rho}(t)} = -\frac{2}{3} \left( \frac{1}{aH} \right)^2 e^{-2\zeta(r)} \left( \zeta''(r) + \frac{2}{r}\zeta'(r) + \frac{1}{2}\zeta'(r)^2 \right) \, .
\end{equation}
A couple of the points to note: First, we see that the density contrast is a second-order quantity in the gradient expansion ($\mathcal{O}(\epsilon^2)$ due to the $1/(aH)^2$ factor), sourced entirely by the non-linear curvature profile $\zeta(r)$.
Second, this explicitly shows that a time-independent curvature profile $\zeta(r)$ sources a growing density mode ($\delta \propto (aH)^{-2} \propto a^2 \propto t$). The collapse begins when this mode enters the horizon ($\epsilon \to 1$).

\section{Connecting the Newtonian and Relativistic Derivations}
\label{sec:NewtonianRelativisticLink}

A keen reader might notice a discrepancy between the Newtonian derivation in Appendix (\ref{app:JeansDetail}) and the Relativistic Gradient Expansion in Sec.~(\ref{sec:DerivationDelta}).
\begin{itemize}
    \item The Newtonian approach yields Eq.~\eqref{eq:Newtonian-delta}.
    \item The Relativistic approach yields a constraint equation for the profile: $\delta \propto \nabla^2 \zeta$.
\end{itemize}
How are these related? We can bridge the gap using the \emph{Separate Universe} intuition (Friedmann equation).

\subsubsection{The Bridge: The Friedmann Constraint}

In the Newtonian limit, we treat gravity as a potential $\phi$ obeying Poisson's equation $\nabla^2 \phi = 4\pi G \delta \rho$.
In the Relativistic Separate Universe limit, we treat the local overdensity as a separate curved FLRW universe. The "potential" is replaced by the 3-dimensional spatial curvature scalar, ${}^{(3)}R$.

Let us start with the Friedmann equation for a local patch with density $\rho$ and spatial curvature ${}^{(3)}R$:
\begin{equation}
H^2 = \frac{8\pi G}{3} \rho - \frac{{}^{(3)}R}{6} \, .
\end{equation}
Rearranging for density $\rho$:
\begin{equation}
\rho = \frac{3H^2}{8\pi G} + \frac{{}^{(3)}R}{16\pi G} \, .
\end{equation}
We assume the background is flat ($\bar{\rho} = \frac{3H^2}{8\pi G}$). The density perturbation $\delta \rho = \rho - \bar{\rho}$ is then entirely determined by the spatial curvature:
\begin{equation}
\delta \rho = \frac{{}^{(3)}R}{16\pi G} \, .
\end{equation}
Dividing by the background density $\bar{\rho} = \frac{3H^2}{8\pi G} = \frac{1}{2} \frac{H^2}{4\pi G}$:
\begin{equation}
\delta \equiv \frac{\delta \rho}{\bar{\rho}} = \frac{{}^{(3)}R / (16\pi G)}{3H^2 / (8\pi G)} = \frac{{}^{(3)}R}{6H^2} \, .
\end{equation}

\subsubsection{Recovering the Gradient Expansion Result}
Now, we calculate the 3-curvature ${}^{(3)}R$ for our perturbed metric $ds^2 \approx -dt^2 + a^2 e^{2\zeta} \delta_{ij} dx^i dx^j$.
To linear order in $\zeta$, the Ricci scalar is:
\begin{equation}
{}^{(3)}R \approx - \frac{4}{a^2} \nabla^2 \zeta \, .
\end{equation}
Substituting this back into the density contrast expression:
\begin{equation}
\delta = \frac{-4 \nabla^2 \zeta / a^2}{6H^2} = -\frac{2}{3} \frac{1}{a^2 H^2} \nabla^2 \zeta \, .
\label{eq:densContra-fin}
\end{equation}
This matches exactly the linearized limit of the full Gradient Expansion result derived in Eq.~\eqref{sec:DerivationDelta}.

\subsubsection{Recovering the Time Evolution (Newtonian Growth)}
Does this static-looking constraint match the time evolution $\delta \propto a^2$?
Recall that on super-horizon scales, the curvature perturbation $\zeta$ is \textbf{conserved} (constant in time).
\begin{itemize}
    \item $\zeta = \text{const}$.
    \item $H^2 \propto a^{-4}$ (in Radiation domination).
    \item Therefore, $\delta \propto \frac{1}{a^2 H^2} \propto \frac{1}{a^2 (a^{-4})} \propto a^2$.
\end{itemize}
This $ \delta \propto a^2$ growth is exactly the growing mode solution of the Newtonian perturbation Eq.~\eqref{eq:Newtonian-delta} in the \ac{RD} era.
Thus we see that the relativistic Gradient Expansion result is simply the non-linear generalization of the Friedmann constraint equation. It tells us that a \emph{constant curvature perturbation} $\zeta$ sources a {growing density contrast $\delta \propto a^2$ on super-horizon scales.

\section{Detailed Derivation of the Collapse Threshold via \texorpdfstring{$\delta$}{delta}}
\label{sec:TopHatDerivation}

To derive the analytic threshold $\delta_{\mathrm{th}} \approx w$, we employ the \emph{Spherical Top-Hat Collapse} model. We model the overdensity as a closed \ac{FLRW} universe embedded within a flat background universe. The reason for this is because we know that in the FLRW closed universe scenario, matter/radiation eventually stops expanding at the turnaround time $t_{\rm max}$ and recollapses. To mimic the formation of a PBH we model the region forming PBH to be 3-space with positive curvature. 
As we will also see the overdense region is related to the geometric quantities.

To go about this, we make a simple approximation~\cite{Carr:1976zz}. The Universe has two distinct regions with the same equation of state $p = w\rho$ is the same in both regions (for radiation, $w=1/3$).:
\begin{enumerate}
    \item \textbf{The Background:} A flat ($k=0$) universe with scale factor $a(t)$ and density $\bar{\rho}(t)$.
    \item \textbf{The Overdensity:} A closed ($k=+1$) universe with scale factor $b(t)$, density $\rho(t)$, and a fixed curvature parameter $K > 0$.
\end{enumerate}
The Friedmann equations for these two regions are:
\begin{align}
    \text{Background:} & \quad H^2 = \left(\frac{\dot{a}}{a}\right)^2 = \frac{8\pi G}{3} \bar{\rho} \, , \label{eq:FriedmannBG} \\
    \text{Overdensity:} & \quad H_b^2 = \left(\frac{\dot{b}}{b}\right)^2 = \frac{8\pi G}{3} \rho - \frac{K}{b^2} \, . \label{eq:FriedmannOD}
\end{align}
Assumption here is that the time coordinate $t$ is the same for both the background and overdense region.


We define the formation epoch $t_i$ as the moment the perturbation $(a_i H_i)$ enters the horizon . At this initial time, we assume the expansion rates are approximately equal ($H_b \approx H$) and the scale factors are matched ($b_i = a_i$).
Using Eq.~\eqref{eq:FriedmannOD}, we can express the curvature $K$ in terms of the initial density contrast $\delta_i = (\rho_i - \bar{\rho}_i)/\bar{\rho}_i$:
\begin{equation}
    H^2 \approx \frac{8\pi G}{3} \bar{\rho}_i (1 + \delta_i) - \frac{K}{a_i^2} \, .
\end{equation}
Subtracting the background Eq.~\eqref{eq:FriedmannBG} ($H^2 = \frac{8\pi G}{3}\bar{\rho}_i$) gives:
\begin{equation}
    0 = \frac{8\pi G}{3} \bar{\rho}_i \delta_i - \frac{K}{a_i^2} \quad \Longrightarrow \quad K = \frac{8\pi G}{3} \bar{\rho}_i a_i^2 \delta_i \, .
\end{equation}
Using $H^2 = \frac{8\pi G}{3}\bar{\rho}$, this simplifies to a crucial geometric relation:
\begin{equation}
    \label{eq:K_delta_relation}
    \frac{K}{a_i^2} = H_i^2 \delta_i \quad \Longrightarrow \quad 
    \delta_i = \frac{K}{(H_i \, a_i)^2} \, .
\end{equation}
Since $H_i a_i$ is related to the horizon scale, we see that the density contrast $\delta$ is the ratio of the curvature scale to the horizon scale squared.

As mentioned above, in the FLRW closed universe scenario, matter/radiation eventually stops expanding at the turnaround time $t_{\rm max}$ and recollapses. This "turnaround" occurs when $\dot{b} = 0$. From Eq.~\eqref{eq:FriedmannOD}, at turnaround time $t_{max}$, we have:
\begin{equation}
    \frac{8\pi G}{3} \rho(t_{max}) = \frac{K}{b^2(t_{max})} \, .
\end{equation}
For a fluid with $w$, density evolves as $\rho \propto b^{-3(1+w)}$. Linking turnaround (max) to initial state ($i$):
\begin{equation}
    \rho_{max} = \rho_i \left( \frac{a_i}{b_{max}} \right)^{3(1+w)} \approx \bar{\rho}_i \left( \frac{a_i}{b_{max}} \right)^{3(1+w)} \, .
\end{equation}
Substituting this into the turnaround condition:
\begin{equation}
    \frac{8\pi G}{3} \bar{\rho}_i \left( \frac{a_i}{b_{max}} \right)^{3(1+w)} = \frac{K}{b_{max}^2} \, .
\end{equation}
Using Eq.~\eqref{eq:K_delta_relation} to replace $K$:
\begin{equation}
    H_i^2 \left( \frac{a_i}{b_{max}} \right)^{3(1+w)} = H_i^2 \delta_i \frac{1}{b_{max}^2} a_i^2 \, .
\end{equation}
Solving for the maximum scale factor $b_{max}$:
\begin{equation}
    \left( \frac{b_{max}}{a_i} \right)^{1+3w} \approx \frac{1}{\delta_i} \, .
\end{equation}
For radiation ($w=1/3$), this gives $b_{max} \approx a_i \delta_i^{-1/2}$.
The time to reach this maximum scale, $t_{max}$, scales roughly as the Hubble time at that scale. Since $a \propto t^{\frac{2}{3(1+w)}}$, we have $t_{max} \approx t_i \delta_i^{-3(1+w)/2}$. For radiation ($w=1/3$), $t_{max} \approx t_i \delta_i^{-1}$.


In calculating the turnaround we have mostly concentrated on the geometric part and we have not included the effect of pressure on the density perturbations. As discussed earlier, the pressure will act as a repellent and will stop the collapse! Specifically, if the pressure forces are too strong, they will disperse the overdensity before it can collapse into a \ac{BH}.
The physical size of the region at turnaround is $R_{phys} \approx b_{max}$.
The \emph{Jeans Length} (sound horizon) at turnaround is the distance a sound wave travels by time $t_{max}$:
\begin{equation}
    R_J \approx c_s t_{max} \approx \frac{c_s}{H_{max}} \, .
\end{equation}
For a successful collapse, gravity must dominate pressure. This requires the physical size of the perturbation to exceed the Jeans length at the moment of turnaround:
\begin{equation}
    \label{eq:JeansCondition}
    b_{max} \gtrsim c_s t_{max} \, .
\end{equation}
Let us check the scaling in the \ac{RD} era ($w=1/3, c_s^2=1/3$).
\begin{itemize}
    \item Physical Size: $b_{max} \propto \delta^{-1/2}$.
    \item Jeans Length: $c_s t_{max} \propto c_s \delta^{-1}$.
\end{itemize}
Substituting these into Eq.~\eqref{eq:JeansCondition}:
\begin{equation}
    \delta^{-1/2} \gtrsim c_s \delta^{-1} \quad \Longrightarrow \quad \delta^{1/2} \gtrsim c_s \, .
\end{equation}
Squaring both sides gives the fundamental threshold condition:
\begin{equation}
    \delta \gtrsim c_s^2 \, .
\end{equation}
Since $c_s^2 = w$ (where $w=p/\rho$), we arrive at the famous approximation~\cite{Carr:1975qj,Carr:1976zz}:
\begin{equation}
\label{eq:deltathreshold}
    \delta_{\mathrm{th}} \approx w = \frac{1}{3} \, .
\end{equation}
This heuristic derivation clarifies that the threshold $\delta \sim 1/3$ arises physically from the competition between the time required to collapse (gravity) and the time required for pressure waves to cross the region (sound speed).

\bibliography{References}
\end{document}